\include{BoxedEPS}

\documentclass[11pt]{amsart}
\usepackage{amsmath,amsthm}

\chardef\bslash=`\\ % p. 424, TeXbook
\makeatletter
\def\verbatim{\interlinepenalty\@M \@verbatim
  \leftskip\@totalleftmargin\advance\leftskip2pc
  \frenchspacing\@vobeyspaces \@xverbatim}
\makeatother
\newtheorem{thm}{Theorem}[section]
\newtheorem{cor}[thm]{Corollary}
\newtheorem{lem}[thm]{Lemma}
\newtheorem{prop}[thm]{Proposition}

\newtheorem{ex}[thm]{Example}

\theoremstyle{definition}
\theoremstyle{remark}
\newtheorem{rem}{Remark}[section]

\numberwithin{equation}{section}

\newcommand{\begeq}{\begin {equation}}

\newcommand{\bs}{\begin {split}}
\newcommand{\es}{\end {split}}
\newcommand{\bp}{\begin {prop}}
\newcommand{\ep}{\end {prop}}
\newcommand{\bt}{\begin {thm}}
\newcommand{\et}{\end {thm}}
\newcommand{\bc}{\begin {cor}}
\newcommand{\ec}{\end {cor}}
\newcommand{\bl}{\begin {lem}}
\newcommand{\el}{\end {lem}}
\newcommand{\bpf}{\begin {proof}}
\newcommand{\epf}{\end {proof}}
\newcommand{\bi}{\begin {itemize}}
\newcommand{\ei}{\end {itemize}}
\newcommand{\ben}{\begin {enumerate}}
\newcommand{\een}{\end {enumerate}}
\newcommand{\brem}{\begin {rem}}
\newcommand{\erem}{\end {rem}}

\newcommand{\ZZ}{{\mathbb Z}}

\newcommand{\RR}{{\mathbb R}}

\newcommand{\Rd}{ {\Bbb R}^d}
\newcommand{\Zd} {{\mathbb Z}^d}

\begin{document}
\bibliographystyle{plain}

\title[Sampling in Reproducing Kernel Spaces] {Sampling and Reconstruction of Signals in a Reproducing Kernel Subspace of $L^p({\Bbb R}^d)$}

%%%%%%%%%%%%%%%%%%%%%%%%%%%%%%%%%%%%%%%%%%%%%%%%%%%%%%%%%%%%%%%%%%%%%%%%
\author{M. Zuhair Nashed }

\address{M. Z. Nashed,\ Department of Mathematics,  University of Central Florida,
Orlando, FL 32816, USA}

%% Note the doubled @@:
\email{znashed@mail.ucf.edu }

\author{Qiyu Sun}
\address{Q. Sun, Department of Mathematics,  University of Central Florida,
Orlando, FL 32816, USA}

\email{qsun@mail.ucf.edu }
%\thanks{Research of the first author was supported in part by NSF grant
%.}
%%%%%%%%%%%%%%%%%%%%%%%%%%%%%%%%%%%%%%%%%%%%%%%%%%%%%%%%%%%%%%%%%%%%%%%%

\date{\today }

%\subjclass{Primary 41A15,42C15, 46A35, 46E15, 46N99, 47B37}
%
%\keywords{Irregular sampling, shift-invariant spaces, wavelets}

%\dedicatory{}

\maketitle
%%%%%%%%%%%%%%%%%%%%%%%%%%%%%%%%%%%%%%%%%%%%%%%%%%%%%%%%%%%%%%%%%%%%%%%%
%%%%%%%%%%%%%%%%%%%%%%%%%%%%%%%%%%%%%%%
%%%%%%%%%%%%%%%%%%%%%%%%%%%%%%%%%
\begin{abstract}
In this paper, we consider sampling and reconstruction of signals in a reproducing kernel  subspace  of $L^p(\Rd), 1\le p\le \infty$,
associated with an idempotent integral operator  whose kernel has certain off-diagonal decay and  regularity.
The space of $p$-integrable non-uniform splines and the shift-invariant spaces generated by finitely many localized functions
are our model examples of such reproducing kernel subspaces of $L^p(\Rd)$.
We show that a signal in  such reproducing kernel subspaces  can be reconstructed in a stable way from its samples taken on a relatively-separated
set with sufficiently small  gap. We also study the exponential convergence, consistency,  and the asymptotic pointwise
 error estimate
of the  iterative approximation-projection
     algorithm
and the iterative frame    algorithm
for reconstructing a signal  in those reproducing kernel spaces from its samples with sufficiently small gap.
\end{abstract}

\tableofcontents

\section{Introduction}

 Sampling  and reconstruction
 is a cornerstone of signal processing.
  The most common form of sampling is the uniform sampling of a bandlimited signal. In this case,
  perfect reconstruction of the signal
  from its uniform samples is possible when the samples are taken at a rate greater than twice the  bandwidth
  \cite{jieee77, sire49}.
  Motivated by the intensive  research activity taking
place around wavelets,
 the paradigm for sampling and reconstructing band-limited signals  has been extended over  the past decade  to signals in
shift-invariant spaces \cite{agsiam01, uieee00}.
 Recently, the  above paradigm  has been further extended  to  representing
signals with finite rate of innovation, which are neither
band-limited nor living in a shift-invariant space \cite{dvbieee07,
mvieee05, sunaicm08, sunsiam, vmbieee02}.
  Here a signal is said to have {\em finite rate of
innovation} if it has finite number of degrees of freedom per unit of time,
that is, if it has requires only a finite number of samples per unit of time to
specify the signal \cite{vmbieee02}.

 \bigskip
 In this paper, we consider  sampling and
 reconstruction of signals in a reproducing kernel subspace of $L^p(\Rd), 1\le p\le \infty$.
 Here and henceforth  $L^p:=L^p(\Rd)$ is the space of all $p$-integrable functions on  the $d$-dimensional Euclidean space $\Rd$
with the  standard norm $\|\cdot\|_{L^p(\Rd)}$, or $\|\cdot\|_p$ for short.
A {\em reproducing kernel subspace of $L^p(\Rd)$} \cite{atams50} is
a closed subspace $V$ of $L^p(\Rd)$ such that
 the evaluation functionals on $V$ are continuous, i.e.,
for any $x\in \Rd$ there exists a positive constant $C_x$ such that
\begin{equation}\label{rks.def.eq}
|f(x)|\le C_x \|f\|_{L^p(\Rd)} \quad \ {\rm for \ all} \ f\in V.
\end{equation}

\bigskip
 Let $1\le p\le \infty$. We say that a bounded linear operator $T$ on $L^p(\Rd)$ is an {\em idempotent} operator
  if it satisfies
\begin{equation}\label{t2=t}
T^2=T.
\end{equation}
Denote by
$V$  the range space  of the idempotent operator $T$ on $L^p(\Rd)$, i.e.,
\begin{equation}\label{rangespace.def} V:=\big\{Tf \ | \ \ f\in L^p(\Rd)\big\}.\end{equation}
We say that the range space $V$ of the idempotent operator $T$ on $L^p(\Rd)$
 is a {\em reproducing kernel space $V$ associated with the idempotent operator $T$ on $L^p(\Rd)$} if
 it is
a reproducing kernel subspace of $L^p(\Rd)$.

\bigskip

A trivial example of idempotent linear operators is the identity operator. In this case, the range space is the whole space
$L^p(\Rd)$  on which  the evaluation functional is not continuous. As pointed out in \cite{nst09}, the whole space $L^2(\Rd)$ is too big
to have  stable sampling and reconstruction  of signals belonging to this space.
So it would be reasonable to have certain additional constraints on the idempotent operator $T$.
In this paper, we further assume that the idempotent operator $T$ is an integral operator
\begin{equation}\label{integraloperator.def}
Tf(x)=\int_{\Rd} K(x,y) f(y)dy, \ \  f\in L^p(\Rd),
\end{equation}
 whose measurable kernel $K$ has certain off-diagonal decay and regularity, namely,
\begin{equation}\label{kernel.assumption1}
\big\|\sup_{z\in \Rd} |K(\cdot+z,z)|\big\|_{L^1(\Rd)}<\infty,
\end{equation}
and
\begin{equation}\label{kernel.assumption2}
\lim_{\delta\to 0} \big\|\sup_{z\in \Rd} |\omega_\delta(K)(\cdot+z,z)|\big\|_{L^1(\Rd)}=0
\end{equation}
\cite{kbook99, sunacha08}.
Here the {\em modulus of continuity} $\omega_\delta(K)$
 of a  kernel function $K$ on $\Rd\times \Rd$   is defined by
 \begin{equation}\label{modulus.def} \omega_\delta(K)(x,y)=\sup_{x', y'\in [-\delta, \delta]^d}
 |K(x+x', y+y')-K(x,y)|.\end{equation}

\bigskip
In this paper, we assume that signals to be  sampled and  represented live in a reproducing kernel space associated with
 an idempotent integral operator  whose  kernel  satisfies \eqref{kernel.assumption1} and \eqref{kernel.assumption2}.
 The reason for this setting is three-fold. First, the range space  of an idempotent integral operator  whose kernel satisfies
\eqref{kernel.assumption1} and \eqref{kernel.assumption2} is a reproducing kernel subspace of $L^p(\Rd)$, see
 Theorem \ref{kernel.tm} in the Appendix. Secondly,   signals in
the range space  of an idempotent integral operator  whose kernel satisfies
\eqref{kernel.assumption1} and \eqref{kernel.assumption2}
 have finite rate of innovation, see
 Theorem \ref{localizedframeforrange.tm} in the Appendix. Thirdly, the common model spaces in sampling theory such as the space of $p$-integrable non-uniform splines of order $n$
 satisfying $n-1$ continuity conditions at each knot \cite{schumakerbook, wbook90} and the finitely-generated shift-invariant space
 with its generators having certain regularity and decay at infinity \cite{agsiam01, uieee00}, are
 the range space of some idempotent integral operators  whose kernels satisfy
\eqref{kernel.assumption1} and \eqref{kernel.assumption2}, see Examples \ref{modelspace1.example} and \ref{modelspace2.example} in the Appendix.

\bigskip
A discrete subset $\Gamma$ of $\Rd$ is said to be {\em relatively-separated} if
\begin{equation}
\label{relativelyseparated.def}
B_\Gamma(\delta):=\sup_{x\in \Rd} \sum_{\gamma\in \Gamma} \chi_{\gamma+[-\delta/2, \delta/2]^d}(x)<\infty
\end{equation}
for some $\delta>0$, while % We remark that $B_\Gamma(\delta)<\infty$ for any $\delta>0$ if it holds for some $\delta>0$.
a positive  number $\delta$ is said to be a {\em gap} of  a  relatively-separated subset $\Gamma$  of $\Rd$   if
  \begin{equation}\label{gap.def}
 A_\Gamma(\delta):=\inf_{x\in \Rd} \sum_{\gamma\in \Gamma} \chi_{\gamma+[-\delta/2, \delta/2]^d}(x)\ge 1
  \end{equation}
  \cite{astjfaa05}.
  Note that the set of all positive numbers $\delta$ with $A_\Gamma(\delta)\ge 1$ is either an interval or an empty set
  because $A_\Gamma(\delta)$ is an increasing function of $\delta>0$. Then for a  relatively-separated subset $\Gamma$  of $\Rd$
    having positive gap, we
 define  the smallest positive number $\delta$ with $A_\Gamma(\delta)\ge 1$ as its {\em maximal gap}.
One may verify that a bi-infinite increasing sequence  $\Lambda=\{\lambda_k\}_{k\in \ZZ}$ of real numbers is
relatively-separated if $\inf_{k\in \ZZ} (\lambda_{k+1}-\lambda_k)>0$, and that it has maximal gap $\sup_{k\in \ZZ} (\lambda_{k+1}-\lambda_k)$
if it is finite.

\bigskip

In this paper, we assume that the sample $Y:=(f(\gamma))_{\gamma\in \Gamma}$ of a signal  $f$
 is taken on a relatively-separated subset $\Gamma$ of $\Rd$ with positive gap.

\bigskip

The samplability is one of most important topics in sampling theory, see for instance
\cite{fgsiam92, gmc92, uieee00}  for band-limited signals, \cite{agsiam01, sunaicm08}
 for signals in a shift-invariant space, \cite{coaam09, fgjfa89, fgmm89, frjfaa05,  gmm91} for signals in a co-orbit space, and
 \cite{hnsappear, mnsaicm03} for signals in  reproducing kernel Hilbert and Banach spaces.
 In this paper, we study the {\em samplability of signals} in a reproducing kernel subspace of $L^p(\Rd)$ associated with an idempotent operator.
Particularly, in  Section \ref{samplability.section},   we   show that  any signal in  a reproducing kernel subspace
 $V$ of $L^p(\Rd)$ associated with an idempotent operator whose kernel satisfies \eqref{kernel.assumption1} and \eqref{kernel.assumption2}
  can be  reconstructed  {\em in a stable way}  from its samples taken on a  relatively-separated set $\Gamma$ with
  sufficiently small gap $\delta$, i.e.,
  there exist positive constants $A$ and $B$ such that
\begin{equation}\label{stability.def.eq}
 A\|f\|_{L^p(\Rd)}\le \|(f(\gamma))_{\gamma\in \Gamma}\|_{\ell^p(\Gamma)}\le B \|f\|_{L^p(\Rd)} \quad \ {\rm for \ all} \  f\in V \end{equation}
(see Theorem \ref{samplable.tm} for the  precise statement).
Here  and henceforth, given a discrete set $\Gamma$,
 $\ell^p:=\ell^p(\Gamma), 1\le p\le \infty$, is the space of all $p$-summable sequences on $\Gamma$ with the standard norm $\|\cdot\|_{\ell^p(\Gamma)}$, or $\|\cdot\|_p$ for short.
\bigskip

In this paper, we then study the linear reconstruction of a
 signal
  from its samples taken on a relatively-separated set with sufficiently small  gap.
The iterative approximation-projection  reconstruction algorithm  is an efficient algorithm
to reconstruct a signal from its samples, which  was
 introduced  in \cite{fgsiam92}
for reconstructing band-limited signals, and was later generalized to signals in shift-invariant
spaces in \cite{afpams98}; see also \cite{astca04, fgjcam08}  and the references therein for various generalizations and applications.
In Section \ref{iterativealgorithm.section} of this paper,
 we  introduce the {\em iterative approximation-projection  reconstruction algorithm} for reconstructing
   a signal in  a reproducing kernel subspace   of $L^p(\Rd)$
    from its samples taken on a relatively-separated set with sufficiently small  gap,
and  study  its exponential convergence, consistency,
 and numerical implementation of the above iterative approximation-projection  algorithm (see Theorem \ref{poorman.tm},  Remark \ref{poorman.tm.noiseremark} and Remark
 \ref{poorman.tm.numericalremark} for details).

\bigskip
Denote the standard action between  functions $f\in L^p(\Rd)$ and $g\in L^{p/(p-1)}(\Rd)$
 by
\begin{equation}\label{lplqaction.def}\langle f, g\rangle=\int_{\Rd} f(x) g(x)dx.\end{equation}
Then  the stability condition \eqref{stability.def.eq} can be interpreted as the $p$-frame property of
 $\{K(\gamma, \cdot)\}_{\gamma\in \Gamma}$ on the space $V$. Here  for a Banach subspace  $V$ of $L^p(\Rd)$,
we say that a family $\Phi=\{\psi_\gamma\}_{\gamma\in \Gamma}$
 of functions  in $L^{p/(p-1)}(\Rd)$ is a {\em $p$-frame for  $V$} \cite{astjfaa01} if there exist positive constants $A$ and $B$ such that
 \begin{equation}\label{pframe.def}
 A\|f\|_{L^p(\Rd)}\le \big\| (\langle f, \psi_\gamma\rangle)_{\gamma\in \Gamma}\big\|_{\ell^p(\Gamma)}\le B \|f\|_{L^p(\Rd)} \quad \ {\rm for \ all} \ f\in V.
 \end{equation}
 Then a natural linear reconstruction algorithm
is the  frame reconstruction algorithm; see \cite{bbook92, yic67} for reconstructing band-limited signals, \cite{agsiam01, aunfao94, cisieee98, lwjfaa96} for reconstructing signals
in shift-invariant spaces, and \cite{nwmcss91} for reconstructing signals in some reproducing kernel Hilbert spaces.
 In Section \ref{framealgorithm.section}, we introduce the {\em preconditioned
  frame algorithm} for reconstructing signals in a reproducing kernel space associated with an
  idempotent integral operator from its samples taken a relatively-separated set $\Gamma$ with sufficiently small gap,
  and study its exponential convergence and consistency (see Theorem \ref{frame.tm} for details).

\bigskip
Reconstructing a function from data corrupted
  by noise and estimating the  reconstruction error are leading problems in sampling theory, however
  they have not been given as much attention; see
  \cite{euieee06, prkieee03,szbams04} for reconstructing bandlimited signals,
  \cite{alsieee08, euieee06} for reconstructing signals in shift-invariant spaces, and \cite{bnsaam09, mvieee05, mvbieee06} for reconstructing signals
  with finite rate of innovations.
  It is observed in \cite{rbhspie05}
  that reconstruction from noisy data may introduce spatially-dependent  noise in the reconstructed signal
  (hence spatial dependent artifacts) that
     are undesirable for sub-pixel signal processing.
     Thus it is desirable to have an accurate error estimate of the reconstructed signal at each  point.
In this paper, we show that
 the  reconstruction
via the approximation-projection reconstruction algorithm   and the frame reconstruction algorithm
is unbiased, and we also  provide an asymptotic estimate of the variance of the error between the reconstruction  from noisy sample of
a signal $f$ via these algorithms and the signal $f$ in a reproducing kernel space,
 see Theorem \ref{error.tm} and Remark \ref{error.rem2}.

\bigskip

The range space $V$ of an idempotent operator $T$ on $L^p(\Rd)$ has various properties. For instance,  it
 is  complementable
and the null space
$N(T):=\{g\in L^p(\Rd)\ | \ Tg=0\}$
is its  algebraic  and   topological complement.
In the appendix,
some properties of the range space of an idempotent  integral operator on $L^p(\Rd)$ whose
kernel satisfies \eqref{kernel.assumption1} and \eqref{kernel.assumption2}
  are established, such as the reproducing kernel property in Theorem \ref{kernel.tm} and the frame property in Theorem \ref{localizedframeforrange.tm}.

\bigskip

\section{Samplability of signals in a reproducing kernel space}
\label{samplability.section}

In this section,   we consider the samplability of signals
 in  a reproducing kernel subspace  $V$ of $L^p(\Rd)$ associated with an idempotent
  integral operator whose kernel satisfies \eqref{kernel.assumption1} and \eqref{kernel.assumption2}, by showing  that any  signal
  in  $V$
  can be  reconstructed in a stable way from its samples taken
on a  relatively-separated set with sufficiently small  gap.

\begin{thm}\label{samplable.tm}
Let $1\le p\le \infty$, $T$ be an idempotent integral operator whose kernel $K$ satisfies \eqref{kernel.assumption1} and \eqref{kernel.assumption2},
$V$ be the reproducing kernel subspace of $L^p(\Rd)$ associated with the operator $T$, and
$\delta_0>0$ be so chosen that
\begin{equation}\label{samplable.tm.eq1}
r_0:=\big\|\sup_{z\in \Rd} |\omega_{\delta_0/2}(K)(\cdot+z,z)|\big\|_{L^1(\Rd)}<1.
\end{equation}
Then
any signal $f$ in $V$ can be reconstructed in a stable way from its samples $f(\gamma), \gamma\in \Gamma$, taken on a relatively-separated subset $\Gamma$
of $\Rd$ with gap $ \delta_0$.
Moreover, 
  \begin{eqnarray} \label{samplable.tm.eq2}
  & &   (1-r_0) \big(\delta_0^{-d} A_\Gamma(\delta_0)\big)^{1/p}\big\|f\big\|_{L^p(\Rd)} \\
   &    \le &
   \big\|(f(\gamma))_{\gamma\in \Gamma}\big\|_{\ell^p(\Gamma)}
\le  (1+r_0) \big(\delta_0^{-d} B_\Gamma(\delta_0)\big)^{1/p}\big\|f\big\|_{L^p(\Rd)}\quad {\rm for \ all} \ f\in V.\nonumber
  \end{eqnarray}
\end{thm}

\bigskip

Now we apply the above samplability result to signals in a shift-invariant space.
 Let
 \begin{equation}{\mathcal W}:=\Big\{f\big|\ \|f\|_{\mathcal W}:=\sum_{k\in \Zd} \sup_{x\in [-1/2, 1/2]^d} |f(x+k)|<\infty\Big\}
 \end{equation}
 be the {\em Wiener amalgam space} \cite{agsiam01, fcjm90}.
  Let $\phi_1, \ldots, \phi_r\in {\mathcal W}$ be   continuous functions  on $\Rd$
  with the property that  $\{\phi_i(\cdot-k): \ 1\le i\le r,  k\in \Zd\}$ is an orthonormal subset of $L^2(\Rd)$.
Then the integral operator $T$ defined by
\begin{equation}\label{sisoperator.def}
Tf(x)=\int_{\Rd} \Big(\sum_{i=1}^r\sum_{k\in \Zd} \phi_i(x-k) \phi_i(y-k)\Big) f(y) dy \quad {\rm for \ all} \ f\in L^2(\Rd)\end{equation}
is an idempotent operator  whose kernel satisfies \eqref{kernel.assumption1} and \eqref{kernel.assumption2}.
This yields  the  samplability of  signals in a  finitely-generated shift-invariant space \cite{afpams98}.

\begin{cor}
Let $\phi_1, \ldots, \phi_r\in {\mathcal W}$ be   continuous functions  on $\Rd$
such that $\{\phi_i(\cdot-k)| \  1\le i\le r, k\in \Zd\}$ is an orthonormal subset of $L^2(\Rd)$.
Define the  finitely-generated shift-invariant space $V_2(\phi_1, \ldots, \phi_r)$ by
\begin{equation}
\label{sis.def}
V_2(\phi_1, \ldots, \phi_r)= \Big\{\sum_{i=1}^r\sum_{k\in \Zd}  c_i(k) \phi_i(\cdot-k)\ \Big |\ \ \sum_{i=1}^r\sum_{k\in \Zd} |c_i(k)|^2<\infty\Big\}.\end{equation}
Then
any signal $f$ in $V_2(\phi_1, \ldots, \phi_r)$ can be reconstructed
in a stable way from its samples $f(\gamma), \gamma\in \Gamma$, taken on a relatively-separated subset $\Gamma$
of $\Rd$ with sufficiently small gap $\delta_0$.
\end{cor}

The following theorem is a slight generalization of Theorem \ref{samplable.tm}.

\begin{thm}\label{samplable.cor2}
Let $1\le p\le \infty$, $T$ be an idempotent integral operator whose kernel $K$ is continuous and satisfies
\begin{equation}
\sup_{x\in \Rd} \|K(x,\cdot)\|_{L^1(\Rd)}+\sup_{y\in \Rd} \|K(\cdot,y)\|_{L^1(\Rd)}<\infty,
\end{equation}
$V$ be the reproducing kernel subspace of $L^p(\Rd)$ associated with the operator $T$, and
$\delta_0>0$ be so chosen that
\begin{eqnarray} \label{samplable.cor2.eq2}
\qquad r_0^\prime &:= &\Big(\sup_{x\in \Rd}\Big\|\sup_{|t|\le \delta_0/2} |K(x+t,\cdot)-K(x,\cdot)|\Big\|_{L^1(\Rd)}\Big)^{1-1/p}
\\
& & \quad \times \Big(\sup_{y\in \Rd}\Big\|\sup_{|t|\le \delta_0/2} |K(\cdot+t, y)-K(\cdot,y)|\Big\|_{L^1(\Rd)}\Big)^{1/p}<1.\nonumber
\end{eqnarray}
Then
any signal $f$ in $V$ can be reconstructed in a stable way from its samples $f(\gamma), \gamma\in \Gamma$, taken on a relatively-separated subset $\Gamma$
of $\Rd$ with gap $ \delta_0$.
\end{thm}

\bigskip
\begin{rem}
{\rm
The conclusion in Theorem \ref{samplable.cor2} is established in \cite[Section 7.5]{frjfaa05} when the kernel $K$ of the
 idempotent operator $T$  satisfies
 \begin{equation}\label{symmetrickernel.assumption}
 K(x,y)=\overline{K(y,x)}.
 \end{equation}
For $p=2$, an idempotent operator $T$ with kernel $K$ satisfying \eqref{symmetrickernel.assumption}
is a projection operator onto a closed subspace of $L^2$. Hence the idempotent operator $T$ with its kernel
satisfying \eqref{symmetrickernel.assumption} is uniquely determined by its range space $V$ onto $L^2$.
The above conclusion on the idempotent operator
 does not hold without the assumption  \eqref{symmetrickernel.assumption} on its kernel.
We leave the above option on the kernel of  idempotent operators  free for  better estimate in
the  gap $\delta_0$ in Theorem \ref{samplable.tm},  and  also for our further study on local exact reconstruction
(c.f. \cite{agjfaa00,   sunaicm09, sunwaicm09} for signals in shift-invariant spaces).
For instance, let us consider  samplability of signals in the linear spline space
$$V_1:=\Big\{ \sum_{k\in \ZZ} c(k) h(x-k)\big| \ \sup_{k\in \ZZ} |c(k)|<\infty\Big\},$$
where $h(x):=\max(1-|x|, 0)$ is the hat function.
 It is well known \cite{agjfaa00}  that  a  signal $f$ in the  linear spline space $V_1$ can be reconstructed in a stable way
  from  its  samples $f(\gamma_k), k\in \ZZ$, with maximal gap $\delta_0:=\sup_{k\in \ZZ} (\gamma_{k+1}-\gamma_k)<1$.
For any  integer $N\ge 1$, define
$$K_N(x,y)=\frac{3N^2}{\sqrt{9N^2-6N}} \sum_{k,l\in \ZZ} h(x-k) h(N(y-l))  \big(\sqrt{9N^2-6N}-3N+1\big)^{|k-l|},$$
and let $T_N$ be the  integral operator with kernel $K_N$.  One may verify that
$T_N, N\ge 1$, are idempotent operators with the same range space $V_1$ and
 the  kernel $K_N$ satisfies \eqref{symmetrickernel.assumption} {\bf only} when $N=1$.
Recalling that $K_N(x-1, y-1)=K_N(x,y)$ and $K_N(-x, -y)=K_N(x,y)$,
we have
\begin{eqnarray*}
 & &   \sup_{x\in \RR} \big\| \sup_{|t|\le \delta_0/2} |K_N(x+t,\cdot)-K_N(x,\cdot)|\big\|_1\\
   & = & \sup_{x\in [0, 1/2]} \big\| \sup_{|t|\le \delta_0/2} |K_N(x+t,\cdot)-K_N(x,\cdot)|\big\|_1\\
    & \le &
\frac{3N^2}{\sqrt{9N^2-6N}} \sum_{s=-\infty}^\infty  \big(3N-1-\sqrt{9N^2-6N}\big)^{|s|}\\
& & \times
     \sup_{x\in [0, 1/2]}
  \big\| \sup_{|t|\le \delta_0/2}
      \sum_{k\in \ZZ} |h(x-k)-h(x+t-k)| h(N(\cdot-k-s))|\big\|_1\\
     & \le &
%\frac{3N^2\delta_0}{2\sqrt{9N^2-6N}} \sum_{s=-\infty}^\infty  \big(3N-1-\sqrt{9N^2-6N}\big)^{|s|}\\
%& & \times
%      \sum_{k=-1, 0, 1} \| h(N(\cdot-k-s))|\big\|_1\\
              \frac{9N\delta_0}{6N-4}.
\end{eqnarray*}
This shows that the  inequality \eqref{samplable.cor2.eq2}  holds for  $K=K_N$ and $p=\infty$
when $\delta_0<\frac{2}{3}-\frac{4}{9N}$. On the other hand, we have
\begin{eqnarray*}
 & &   \sup_{x\in \RR} \big\| \sup_{|t|\le \delta_0/2} |K_1(x+t,\cdot)-K_1(x,\cdot)|\big\|_1
    \\
    & \ge &  \|K_1(\delta_0/2,\cdot)-K_1(0,\cdot)\|_1 =
% & = &
% \frac{3 N^2\delta_0}{2\sqrt{9N^2-6}}\\
% &   & \times
%\big\|\sum_{l\in \ZZ}  \big(\sqrt{9N^2-6}-3N^2+1)^{|l|}-(\sqrt{9N^2-6}-3N^2+1)^{|1-l|}\big)h(\cdot-l)\big\|_1\\
%&  = &
\frac{(9-\sqrt{3})\delta_0}{4},\end{eqnarray*}
which implies that
 the  inequality \eqref{samplable.cor2.eq2} does not  hold for  $K=K_1$ and $p=\infty$
when $\delta_0\ge \frac{4}{(9-\sqrt{3})}\approx 0.5504$ and so the theorem does not apply.
}
\end{rem}

\bigskip
We  conclude this section by providing proofs of Theorems \ref{samplable.tm} and
\ref{samplable.cor2}.
To prove Theorem \ref{samplable.tm}, we need a technical lemma.

\begin{lem}\label{stability.lem} Let $1\le p\le  \infty$, $\delta_0\in (0, \infty)$, $r\in (0,1)$,
% $w$ be a   weight (i.e., a positive function)  on $\Rd$,
and $\Gamma$ be a  discrete subset of $\Rd$ with the property that
\begin{equation}\label{stability.lem.eq0}
1\le A_\Gamma(\delta_0)\le B_\Gamma(\delta_0)<\infty.\end{equation}
 Assume that  $f\in L^p(\Rd)$ satisfies
\begin{equation}\label{stability.lem.eq1}
\|\omega_{\delta_0/2}(f)\|_{L^p(\Rd)}\le r\|f\|_{L^p(\Rd)},
\end{equation}
and  $U:=\{u_\gamma\}_{\gamma\in \Gamma}$ is a
 bounded uniform partition of unity (BUPU)  associated with the  covering
 $\{\gamma+[-\delta_0/2,  \delta_0/2]^d\}_{\gamma\in \Gamma}$ of $\Rd$, i.e.,
\begin{equation}\label{stability.lem.eq1b}
\left\{\begin{array}{l} 0\le u_\gamma(x)\le 1 \ {\rm for \ all} \ x\in \Rd \ {\rm and } \  \gamma\in \Gamma,\\
  u_\gamma \ {\rm is\ supported\ in}\ \gamma+[-\delta_0/2, \delta_0/2]^d \ {\rm  for \  each} \  \gamma\in \Gamma, {\rm and} \\
  \sum_{\gamma\in \Gamma} u_\gamma(x)\equiv 1 \ {\rm for \ all} \ x\in \Rd. \end{array}\right.
\end{equation}
Then
\begin{equation}\label{stability.lem.eq2}
(1-r)\|f\|_{L^p(\Rd)}\le  \big\| \big(f(\gamma)\|u_\gamma\|_{L^1(\Rd)}^{1/p}\big)_{\gamma\in \Gamma}\big\|_{\ell^p(\Gamma)}\le (1+r) \|f\|_{L^p(\Rd)}.
\end{equation}
\end{lem}

\begin{proof} By the definition of the modulus of continuity,
\begin{equation}
\label{stability.lem.pf.eq1}
|f(x)|-|\omega_{\delta_0/2}(f)(x)|\le |f(\gamma)|\le |f(x)|+|\omega_{\delta_0/2}(f)(x)|
\end{equation}
for all  $   x\in \gamma+[-\delta_0/2, \delta_0/2]^d$ and $\gamma\in \Gamma$.  This together with
 \eqref{stability.lem.eq0} and \eqref{stability.lem.eq1}  proves  \eqref{stability.lem.eq2}.

For  $1\le p<\infty$, it follows from
\eqref{stability.lem.eq1}, \eqref{stability.lem.eq1b}, and \eqref{stability.lem.pf.eq1} that
\begin{eqnarray*}
\|f\|_{p} %(\int_{\Rd} |f(x)|^p w(x) dx)^{1/p}
& =  &  \Big(\sum_{\gamma\in \Gamma} \int_{\Rd} |f(x)|^p u_\gamma(x) dx\Big)^{1/p}\\
%& \le &
%\Big(\sum_{\gamma\in \Gamma} \|u_\gamma\|_{p, w}  |f(\gamma)|^p\Big)^{1/p}\\
%& & +
%\Big(\sum_{\gamma\in \Gamma} \int_{\Rd} |f(x)-f(\gamma)|^p u_\gamma(x) dx\Big)^{1/p}\\
& \le   &  \Big(\sum_{\gamma\in \Gamma}\int_{\Rd}  |f(\gamma)|^p u_\gamma(x)dx\Big)^{1/p}+
\Big(\sum_{\gamma\in \Gamma} \int_{\Rd} |\omega_{\delta_0/2}(f)(x)|^p u_\gamma(x)  dx\Big)^{1/p}\\
%& \le & \Big(\sum_{\gamma\in \Gamma}\|u_\gamma\|_{1}  |f(\gamma)|^p\Big)^{1/p}+\|\omega_{\delta_0}(f)\|_{p}
%\\
& \le &
\Big(\sum_{\gamma\in \Gamma}|f(\gamma)|^p\|u_\gamma\|_{1} \Big)^{1/p}+r\|f\|_{p},
\end{eqnarray*}
and
\begin{eqnarray*}
 \Big(\sum_{\gamma\in \Gamma}   |f(\gamma)|^p\|u_\gamma\|_{1}\Big)^{1/p} %\\
& \le &
 \Big(\sum_{\gamma\in \Gamma} \int_{\Rd} \big||f(x)|+\omega_{\delta_0/2} (f)(x)\big|^p  u_\gamma(x) dx\Big)^{1/p}\\%+
% \Big(\sum_{\gamma\in \Gamma} \int_{\Rd}   u_\gamma(x) w(x) dx\Big)^{1/p}\\
 & \le & (1+r) \|f\|_p.
\end{eqnarray*}
Then   \eqref{stability.lem.eq2}  for $1\le p<\infty$ is proved.
%how to have great form for $p=\infty$?
\end{proof}

\begin{rem}\label{unitpartition.rem}
 Two popular  examples of  bounded uniform partitions of unity (BUPU)  associated with the  covering
 $\{\gamma+[-\delta_0/2,  \delta_0/2]^d\}_{\gamma\in \Gamma}$ of $\Rd$
 are  given by
 \begin{equation}\label{unitpartition.rem.eq1}
 u_\gamma(x)=\frac{\chi_{\gamma+[-\delta_0/2, \delta_0/2]^d }(x)} {\sum_{\gamma'\in \Gamma} \chi_{\gamma'+[-\delta_0/2, \delta_0/2]^d}(x)}, \ \gamma\in \Gamma,\end{equation}
and
\begin{equation}\label{unitpartition.rem.eq2}
u_\gamma(x)=\chi_{_{V_\gamma}}(x), \ \gamma\in \Gamma,\end{equation}
where $V_\gamma$ is the  Voronoi polygon whose interior
consists of all points in  $\Rd$ being closer to $\gamma$ than any other point
$\gamma'\in \Gamma$.
\end{rem}

Given a  continuously differentiable function $f$ on the real line, its modulus  of continuity $\omega_\delta(f)(x)$ is dominated by
the integral of its derivative $f'$ on $x+[-\delta, \delta]$, i.e.,
$$\omega_{\delta}(f)(x)\le \int_{-\delta}^{\delta}|f'(x+t)|dt \quad \ {\rm for \ all} \  x\in \RR.$$
Then the following result (which is well known for band-limited signals \cite{fgsiam92}) follows easily from Lemma \ref{stability.lem}.

\begin{cor}\label{stability.cor} Let   $1\le p\le \infty$, $f$ be a  time signal satisfying
\begin{equation}
\|f'\|_{L^p(\RR)}\le  B_0\|f\|_{L^p(\RR)}
\end{equation}
for some positive constant $B_0$, and
$\Gamma=\{\gamma_k\}_{k\in \ZZ}$ be a relatively-separated subset of $\RR$ with maximal gap $\delta_0<1/B_0$.
Then there exists a  positive constant $C$ (that depends on $B_0, B_\Gamma(\delta_0)$ and $A_\Gamma(\delta_0)$ only) such that
\begin{equation}\label{stability.cor.eq2}
C^{-1}\|f\|_{L^p(\RR)}\le \big\| \big(f(\gamma)\|u_\gamma\|_{L^1(\Rd)}^{1/p}\big)_{\gamma\in \Gamma}\big\|_{\ell^p(\Gamma)}\le C \|f\|_{L^p(\RR)}.
\end{equation}
\end{cor}

Now we  prove Theorem \ref{samplable.tm}.

\begin{proof} [Proof of Theorem \ref{samplable.tm}]
For any $f\in V$, 
\begin{eqnarray} \label{samplable.tm.pf.eq4}
\|\omega_{\delta_0/2}(f)\|_p & = & \|\omega_{\delta_0/2}(Tf)\|_p\le  %\|\psi\|_1\|\omega_\delta(f)\|_p\le
 \Big\| \int_{\Rd} \omega_{\delta_0/2}(K) (\cdot, y)|f(y)|dy\Big\|_p\\
  & \le &   \big\|\sup_{z\in \Rd} |\omega_{\delta_0/2}(K)(\cdot+z,z)|\big\|_{1}\|f\|_p=r_0\|f\|_p.\nonumber
%
% \nonumber\\
%& \le & \|h_\delta\|_{\mathcal W}\|f\|_p\le \|\psi\|_1 \|h_\delta\|_{\mathcal W}(A_{\psi})^{-1}
%\|\psi*f\|_p. %\\
%& \le & r \|\psi*f\|_p
\end{eqnarray}
%Let $\delta_0>0$ be so chosen that
%\begin{equation} \label{samplable.tm.pf.eq5}
%%\|\psi\|_1
%\|h_{\delta_0/2}\|_{\mathcal W}<1.
%%<A_{\psi}.
%\end{equation}
%The existence of such a positive number $\delta_0$ follows from \eqref{samplable.tm.pf.eq3}.
For any  discrete set $\Gamma$ with $1\le A_\Gamma(\delta_0)\le B_\Gamma(\delta_0)<\infty$,  we
 define $\{u_\gamma\}_{\gamma\in \Gamma}$ as in \eqref{unitpartition.rem.eq1}.
Then \begin{equation}\label{samplable.tm.pf.eq6} \frac{\delta_0^d}{B_\Gamma(\delta_0)}\le
\|u_\gamma\|_1\le \frac{\delta_0^d}{A_\Gamma(\delta_0)} \quad {\rm for \ all} \ \gamma\in \Gamma.
\end{equation}
  From \eqref{samplable.tm.eq1},
\eqref{samplable.tm.pf.eq4}  and Lemma \ref{stability.lem},  we obtain the estimates in
\eqref{samplable.tm.eq2}  for $p=\infty$.
%\begin{eqnarray*}
%\sup_{\gamma\in \Gamma} |f(\gamma)|
%\le  %\frac{A_\psi+\|\psi\|_1 \|h_{\delta_0}\|_{\mathcal W}}{A_\psi}
%%\|\psi*f\|_\infty \le B_\psi
%\|f\|_\infty,
%\end{eqnarray*}
%and
%\begin{eqnarray*}
%\sup_{\gamma\in \Gamma} |f(\gamma)|
%\ge   (1- \|h_{\delta_0/2}\|_{\mathcal W}) %\frac{A_\psi-\|\psi\|_1 \|h_{\delta_0}\|_{\mathcal W}}{A_\psi}
%\|f\|_\infty, %\ge (A_\psi-\|\psi\|_1 \|h_{\delta_0}\|_{\mathcal W}) \|f\|_\infty,
%\end{eqnarray*}
%which prove \eqref{samplable.tm.eq2}.
On the other hand, from % \eqref{samplable.tm.pf.eq1},
 \eqref{samplable.tm.eq1}, \eqref{samplable.tm.pf.eq4},
  \eqref{samplable.tm.pf.eq6} and Lemma \ref{stability.lem},  we get the following estimate for $1\le p<\infty$:
   \begin{eqnarray*}
 %& &
  \Big(\sum_{\gamma\in \Gamma} |f(\gamma)|^p\Big)^{1/p}%\\
 & \le & (\delta_0^{-d} B_\Gamma(\delta_0))^{1/p}  \Big(\sum_{\gamma\in \Gamma} |f(\gamma)|^p \|u_\gamma\|_1 \Big)^{1/p}\\
 & \le &  (\delta_0^{-d} B_\Gamma(\delta_0))^{1/p} (1+r_0)\|f\|_p
 %\frac{ (B_\Gamma)^{1/p} (A_\psi+\|\psi\|_1 \|h_{\delta_0}\|_{\mathcal W}) B_\psi}{(2\delta_0)^{d/p} A_\psi}\|f\|_p
 \end{eqnarray*}
 and
  \begin{eqnarray*}
 %& &
  \Big(\sum_{\gamma\in \Gamma} |f(\gamma)|^p\Big)^{1/p}%\\
 & \ge & (\delta_0^{-d} A_\Gamma(\delta_0))^{1/p}  \Big(\sum_{\gamma\in \Gamma} |f(\gamma)|^p \|u_\gamma\|_1 \Big)^{1/p}\\
 & \ge & (\delta_0^{-d} A_\Gamma(\delta_0))^{1/p}  (1-r_0)\|f\|_p.
  %\frac{ (A_\Gamma)^{1/p} (A_\psi-\|\psi\|_1 \|h_{\delta_0}\|_{\mathcal W}) }{(2\delta_0)^{d/p} }\|f\|_p,
 \end{eqnarray*}
This   proves  \eqref{samplable.tm.eq2} for $1\le p<\infty$.
\end{proof}

\begin{proof} [Proof of Theorem \ref{samplable.cor2}] Similar argument used in the proof of Theorem \ref{samplable.tm} can be applied to
prove Theorem \ref{samplable.cor2}.  We leave the detailed proof for the interested readers.
%We omit the details of the proof here.
\end{proof}
\bigskip

\section{Iterative approximation-projection  reconstruction algorithm}
\label{iterativealgorithm.section}

In this section,   we  show that  signals in  a reproducing kernel subspace  of $L^p(\Rd)$  associated with an idempotent integral operator
 can be reconstructed,
via an  iterative approximation-projection reconstruction algorithm, from its samples taken
on a  relatively-separated set with sufficiently small  gap.

\begin{thm}\label{poorman.tm}
Let $1\le p\le \infty$, $T$ be an idempotent integral operator whose kernel $K$  satisfies \eqref{kernel.assumption1} and \eqref{kernel.assumption2},
$V$ be the reproducing kernel subspace of $L^p(\Rd)$ associated with the operator $T$, and
$\delta_0>0$ be so chosen that \eqref{samplable.tm.eq1} holds.
Set
$$r_0:=\big\|\sup_{z\in \Rd} |\omega_{\delta_0/2}(K)(\cdot+z,z)|\big\|_{L^1(\Rd)}.$$
 Then for any relatively-separated subset $\Gamma$ with  gap $\delta_0$ and $c_0=(c_0(\gamma))_{\gamma\in \Gamma}\in \ell^p(\Gamma)$,
the sequence  $\{f_n\}_{n=0}^\infty$ of  signals in $V$
defined by
\begin{equation}\label{poorman.tm.eq2}
\left\{\begin{array}{l}
f_0(x)= \sum_{\gamma\in \Gamma} c_0(\gamma)\  Tu_\gamma(x), \\
f_n(x)= f_0(x)+ f_{n-1}(x)-  \sum_{\gamma\in \Gamma} f_{n-1}(\gamma)\  Tu_\gamma(x) \quad  {\rm for } \ n\ge 1,
\end{array}\right.
\end{equation}
converges  exponentially, precisely
  \begin{equation}\label{poorman.tm.eq3}
\|f_n-f_\infty\|_{L^p(\Rd)}\le \|T\|\|f_0\|_{L^p(\Rd)} r_0^{n+1} /(1-r_0) \quad \ {\rm for\ some }\  f_\infty\in V,
%\big(\big\|\sup_{z\in \Rd}\omega_{\delta_0/2}(K)(\cdot+z, z)\big\|_1)^n
%\le  C \big\|\sup_{z\in \Rd} |K(\cdot+z, z)|\big\|_1  (\sup_{\gamma\in \Gamma} \|u_\gamma\|_1)^{1/p} \|c_0\|_p  r_0^n
  \end{equation}
where $U:=\{u_\gamma\}_{\gamma\in \Gamma}$ is a BUPU in \eqref{stability.lem.eq1b}.
The sample of the limit signal $f_\infty$ and the given  initial data $c_0$ are related by
\begin{equation}\label{poorman.tm.eq4}
\sum_{\gamma\in \Gamma} \big(c_0(\gamma)-f_\infty(\gamma)\big)  Tu_\gamma(x)\equiv 0.
\end{equation}
Furthermore the iterative algorithm \eqref{poorman.tm.eq2} is consistent, i.e.,
if the given initial data $c_0=(g(\gamma))_{\gamma\in \Gamma}$ is obtained by  sampling a  signal $g\in V$
 then the sequence  $\{f_n\}_{n=0}^\infty$ in  the iterative algorithm \eqref{poorman.tm.eq2}
converges to $g$.
\end{thm}

\begin{proof} %[Proof of Theorem \ref{poorman.tm}]\
Define a  bounded operator $Q_{\Gamma, U}$  on $L^p$ by
\begin{eqnarray}\label{poorman.tm.pf1.eq1}\quad
Q_{\Gamma, U} f(x) & := & \sum_{\gamma\in \Gamma} (Tf)(\gamma) u_{\gamma}(x)-(Tf)(x)\\
 & = & \int_{\Rd} \Big(\sum_{\gamma\in \Gamma} u_\gamma(x) K(\gamma, y)-K(x,y)\Big) f(y) dy,  \ f\in L^p.\nonumber
\end{eqnarray}
Then 
\begin{equation}\label{poorman.tm.pf1.eq2}
Q_{\Gamma,U} T= Q_{\Gamma, U}
\end{equation}
by   \eqref{t2=t},
and
\begin{equation}\label{poorman.tm.pf1.eq3}
\|Q_{\Gamma,U} f\|_p\le r_0 \|f\|_p \quad  {\rm for \ all} \  \ f\in L^p
\end{equation}
 by  the following estimate for the integral kernel of the operator $Q_{\Gamma, U}$:
 \begin{equation}\label{poorman.tm.pf1.eq3b}
 \Big|\sum_{\gamma\in \Gamma} u_\gamma(x) K(\gamma, y)-K(x,y)\Big|\le  %|\omega_{\delta_0/2}(K)(x, y)|\le
\sup_{z'\in \Rd} \big|\omega_{\delta_0/2}(K)(x-y+z', z')\big|.\end{equation}

Define the approximation-projection operator $P_{\Gamma, U}$ by
\begin{equation}\label{poorman.tm.pf1.eq4}
 P_{\Gamma, U}= T Q_{\Gamma, U}+T.\end{equation}
 Then it follows from \eqref{t2=t}, \eqref{poorman.tm.pf1.eq2} and \eqref{poorman.tm.pf1.eq3} that
\begin{equation}\label{poorman.tm.pf1.eq5}
P_{\Gamma, U} T=T P_{\Gamma, U} = P_{\Gamma, U},
\end{equation}
\begin{equation} \label{poorman.tm.pf1.eq5b}
(T-P_{\Gamma, U})^n= (-1)^n T Q_{\Gamma, U}^n\quad  {\rm for \ all}\  n\ge 1,
\end{equation}
and
\begin{eqnarray}\label{poorman.tm.pf1.eq6}
 \|(T-P_{\Gamma,U})^n\| \le   \|T\| r_0^n \quad  {\rm for \ all}\  n\ge 1.
\end{eqnarray}

By   \eqref{poorman.tm.eq2}, \eqref{poorman.tm.pf1.eq1} and \eqref{poorman.tm.pf1.eq4},
\begin{eqnarray}\label{poorman.tm.pf1.eq7}
f_{n+1}-f_n & = &  (T- P_{\Gamma, U}) (f_n-f_{n-1})\\ & = &  \cdots\nonumber\\
 & =  & (T-P_{\Gamma, U})^{n} (f_1-f_0) \nonumber\\ &= &  (T-P_{\Gamma,U})^{n+1} f_0, %\nonumber\\
%  & = & (-1)^{n+1} T Q_{\Gamma,U}^{n+1} f_0,
\ \ n\ge 0.\nonumber
\end{eqnarray}
This together with \eqref{poorman.tm.pf1.eq6} proves  the exponential  convergence of $f_n, n\ge 0$, and
the estimate \eqref{poorman.tm.eq3}.

\bigskip

The equation \eqref{poorman.tm.eq4} follows easily by taking limit  on both sides of
\eqref{poorman.tm.eq2} and applying \eqref{samplable.tm.eq2}.

\bigskip

Define
\begin{equation}\label{poorman.tm.pf1.eq9}
R_{\rm AP}:= T+\sum_{n=1}^\infty (T-P_{\Gamma,U})^n.
\end{equation}
Then it follows from \eqref{poorman.tm.pf1.eq5} and \eqref{poorman.tm.pf1.eq6} that
$R_{\rm AP}$ is a bounded operator on $L^p$ and
a pseudo-inverse of the operator $P_{T,U}$, i.e.,
\begin{equation} \label{poorman.tm.pf1.eq10}
 R_{\rm AP}P_{\Gamma, U}=P_{\Gamma, U} R_{\rm AP}=T,
\end{equation}
and moreover it satisfies
$$R_{\rm AP}T=TR_{\rm AP}=R_{\rm AP}.$$
Applying \eqref{poorman.tm.pf1.eq7} iteratively leads to
\begin{equation}\label{poorman.tm.pf1.eq11}
f_n= \Big(T+\sum_{k=1}^n (T-P_{\Gamma,U})^k\Big) f_0 \quad {\rm for \ all} \ n\ge 1,
\end{equation}
which  together with \eqref{poorman.tm.pf1.eq9} implies that
\begin{equation}\label{poorman.tm.pf1.eq12}
f_\infty= \lim_{n\to \infty} f_n=R_{\rm AP} f_0.
\end{equation}
In the case that  the initial data $c_0$ is the sample  of a signal $g\in V$,  the initial signal $f_0$ in the
iterative algorithm \eqref{poorman.tm.eq2}  and the  signal $g$ are related by
\begin{equation}\label{poorman.tm.pf1.eq13}
f_0= P_{\Gamma,U} g.
\end{equation}
Combining \eqref{poorman.tm.pf1.eq10}, \eqref{poorman.tm.pf1.eq12} and \eqref{poorman.tm.pf1.eq13} proves the consistency of
the  iterative algorithm \eqref{poorman.tm.eq2}.
\end{proof}

From the proof of Theorem \ref{poorman.tm}, we have the following result for the operator $R_{\rm AP}$ in  \eqref{poorman.tm.pf1.eq9}.

\begin{cor}\label{apkernel.cor}
Let $1\le p\le \infty$, $T$ be an idempotent integral operator whose kernel $K$  satisfies \eqref{kernel.assumption1} and \eqref{kernel.assumption2},
$V$ be the reproducing kernel subspace of $L^p(\Rd)$ associated with the operator $T$,
$\delta_0>0$ be so chosen that \eqref{samplable.tm.eq1} holds,
$\Gamma$ be a relatively-separated subset  with  gap  $\delta_0$,
$U:=\{u_\gamma\}_{\gamma\in \Gamma}$ is a BUPU in \eqref{stability.lem.eq1b}, and
  $R_{\rm AP}$ be as in \eqref{poorman.tm.pf1.eq9}. Then $R_{\rm AP}$ is a bounded integral operator
 on $L^p(\Rd)$
 and its kernel $K_{\rm AP}$ satisfies \eqref{kernel.assumption1},
 \eqref{kernel.assumption2}, and
 \begin{equation}\label{apkernel.rem.eq1}
K_{\rm AP}(x,y)=\int_{\Rd}\int_{\Rd} K(x,z_1)K_{\rm AP}(z_1, z_2) K(z_2, y) dz_1dz_2\quad \ {\rm for \ all}\  x, y\in \Rd. \end{equation}
\end{cor}

\begin{rem}\label{poorman.tm.noiseremark}
If  the initial sample $c_0$ in the iterative approximation-projection  reconstruction algorithm
\eqref{poorman.tm.eq2} is the corrupted sample of a signal $g\in V$, i.e.,
$$c_0=(g(\gamma)+\epsilon(\gamma))_{\gamma\in \Gamma}$$ for some noise $\epsilon=(\epsilon(\gamma))_{\gamma\in \Gamma}$, then the $L^p$ norm of the original signal $g$ and
 the recovered signal $f_\infty$ via the iterative approximation-projection  reconstruction algorithm
\eqref{poorman.tm.eq2} is bounded by  the $\ell^p$ norm of the noise $\epsilon$. More precisely,
 from  \eqref{poorman.tm.pf1.eq6} and \eqref{poorman.tm.pf1.eq7} we obtain
\begin{eqnarray}\label{poorman.tm.rem1.eq2}
    & & \|f_n-g\|_p\\
    & \le  & \sum_{k=n+1}^\infty  \| T Q_{\Gamma, U}^k (f_0-h_0)\|_p+\sum_{k=0}^n  \|T Q_{\Gamma, U}^k  h_0\|_p\nonumber \\
    & \le &\|T\| \sum_{k=n+1}^\infty r_0^k \|f_0-h_0\|_p+\|T\| \sum_{k=0}^n r_0^k \|h_0\|_p\nonumber\\
     & \le & \|T\| (1-r_0)^{-1}  ( \|f_0\|_p r_0^{n+1}+\|h_0\|_p)\nonumber \\
     & \le &\|T\|^2
     (1-r_0)^{-1}   \big(\sup_{\gamma\in \Gamma} \|u_\gamma\|_1\big)^{1/p}
     (\|c_0\|_p r_0^{n+1}+ \|\epsilon\|_p)   \nonumber
\end{eqnarray}
and
\begin{eqnarray}\label{poorman.tm.rem1.eq1}
 \|f_\infty-g\|_p   &\le& \|T\|  (1-r_0)^{-1} \|h_0\|_p\\
 & \le &  \|T\|^2(1-r_0)^{-1}  \big(\sup_{\gamma\in \Gamma} \|u_\gamma\|_1\big)^{1/p} \|\epsilon\|_p,\nonumber
\end{eqnarray}
where  $h_0=\sum_{\gamma\in \Gamma}\epsilon(\gamma)  Tu_\gamma$ and $f_n, n\ge 0$, are given in the
 approximation-projection  reconstruction algorithm \eqref{poorman.tm.eq2}.
Define the sample-to-noise ratio  in the logarithmic decibel scale,
 a term for the power ratio between a sample and the background noise,    by
\begin{equation}\label{poorman.tm.rem1.eq3}
{\rm SNR} ({\rm dB}) =20 \log_{10} \frac{\|c_0\|_p} {\|\epsilon\|_p}.
\end{equation}
%(\sum_{\gamma\in \Gamma} |\epsilon(\gamma)|^p)}{(\sum_{\gamma\in \Gamma} |c_0(\gamma)|^p)^{1/p})}
The estimate in \eqref{poorman.tm.rem1.eq2} suggests that
the stopping step $n_0$ for the iterative approximation-projection  reconstruction algorithm \eqref{poorman.tm.eq2} is
\begin{equation}\label{poorman.tm.rem1.eq4}
n_0=\left[\frac{{\rm  SNR (dB)}}{20 \ln_{10} (1/r_0)}\right],
% \frac{(\sum_{\gamma\in \Gamma} |\epsilon(\gamma)|^p)}{(\sum_{\gamma\in \Gamma} |c_0(\gamma)|^p)^{1/p})}
\end{equation}
where $[x]$ denotes the integral part of a real number $x$.
In this case,
\begin{equation}\label{poorman.tm.rem1.eq5}
\|f_{n_0}-g\|_p\le
  2
\|T\|^2
     (1-r_0)^{-1}   \big(\sup_{\gamma\in \Gamma} \|u_\gamma\|_1\big)^{1/p}
\|\epsilon\|_p,\end{equation}
and   the error between the resulting signal $f_{n_0}$
  and the original signal $g$ is about twice the
 error due to the noise in  the initial sample data.
 \end{rem}

\begin{rem}\label{poorman.tm.numericalremark}
 Given the initial data  $c_0=(c_0(\gamma))_{\gamma\in \Gamma}$, define
\begin{equation}\label{poorman.tm.rem2.eq1}
F_n=(f_n(\gamma))_{\gamma\in \Gamma}, \quad  n\ge 0,
\end{equation}
and
\begin{equation}\label{poorman.tm.rem2.eq2}
A= \big( (Tu_{\gamma'})(\gamma)\big)_{\gamma, \gamma'\in \Gamma},
\end{equation}
where  $f_n, n\ge 0$, is given in the iterative approximation-projection  reconstruction algorithm \eqref{poorman.tm.eq2}.
This leads to
the {\em discrete version of
the  iterative approximation-projection  reconstruction algorithm \eqref{poorman.tm.eq2}}:
\begin{equation}\label{poorman.tm.rem2.eq3}
  \left\{\begin{array}{l} F_0=Ac_0,\\
  F_n= F_0+ (I-A)F_{n-1}, \quad   n\ge 1.
  \end{array}\right.
\end{equation}

{\bf Exponential convergence}:\quad Now let us consider the exponential convergence of the sequence $F_n, n\ge 0$,
when \eqref{kernel.assumption1}, \eqref{kernel.assumption2} and \eqref{samplable.tm.eq1} hold.
By \eqref{poorman.tm.rem2.eq3}, we have
\begin{equation}\label{poorman.tm.rem2.eq4}
F_{n}-F_{n-1}= (I-A)^{n} F_0= (I-A)^{n} A c_0,  \quad \ n\ge 1.
\end{equation}
Define
\begin{equation}\label{poorman.tm.rem2.eq5}
\|c\|_{p,U}= \Big\|\sum_{\gamma\in \Gamma} |c(\gamma)| u_\gamma\Big\|_p \quad {\rm for }\ c=(c(\gamma))_{\gamma\in \Gamma},
\end{equation}
where $1\le p\le \infty$.
For $c=(c(\gamma))_{\gamma\in \Gamma}$ with $\|c\|_{p,U}<\infty$,
 write $(I-A)^n Ac=(d_n(\gamma))_{\gamma\in \Gamma}$ and define
$c_{\Gamma,U} (x)=\sum_{\gamma\in \Gamma}  c(\gamma) u_{\gamma}(x)$.
Similar to the equation \eqref{poorman.tm.pf1.eq6}
 we have
\begin{equation}\label{poorman.tm.rem2.eq6}
d_n(\gamma) =(-1)^{n} (TQ_{\Gamma,U}^n c_{\Gamma,U} )(\gamma).
%\int_{\Rd}\cdots \int_{\Rd}  K(\gamma, y_1) K_{\Gamma, U}(y_1, y_2) \cdots
%K_{\Gamma, U}(y_n,z) c_{\Gamma,U} (z) dz d y_n \cdots dy_1.
\end{equation}
This together with \eqref{poorman.tm.pf1.eq3} implies that
\begin{eqnarray}\label{poorman.tm.rem2.eq7}
 &  & \|(I-A)^n Ac\|_{p,U} \\
& \le &
\Big\| \sum_{\gamma\in \Gamma}  u_\gamma(\cdot) \int_{\Rd}  |K(\gamma, z)|  | (Q_{\Gamma,U}^n c_{\Gamma,U})(z)| dz\Big\|_p\nonumber\\
&\le &
\Big\| \int_{\Rd}  \big(|K(\cdot, z)|+ |\omega_{\delta_0/2}(K)(\cdot, z)|\big)  | (Q_{\Gamma,U}^n c_{\Gamma,U})(z)| dz\Big\|_p\nonumber\\
&\le &  C_0 r_0^n \|c\|_{p,U}\nonumber
%  &\le  &
%  \Big\| \int_{\Rd}\cdots \int_{\Rd}  \big( h(\cdot-y_1)+h_{\delta_0/2}(\cdot-y_1)\big)\nonumber \\
%   & & \quad \times  |h_{\delta_0/2}(y_1-y_2)| \cdots
%|h_{\delta_0/2}(y_n-z)|  |c_{\Gamma,U} (z)| dz d y_n \cdots dy_1\Big\|_p\nonumber\\
% & \le &
% (\|h\|_1+\|h_{\delta_0/2}\|_1) \| Q_{\Gamma,U}^n c_{\Gamma,U}\|_p\le \|h_{\delta_0/2}\|_1^n \|c\|_{p,U}.\nonumber
 \end{eqnarray}
where
\begin{equation}
C_0=
\big \|\sup_{z\in \Rd} |K(\cdot+z,z)|\big\|_1+\big \|\sup_{z\in \Rd} \omega_{\delta_0/2}(K)(\cdot+z,z)\big\|_1.
 \end{equation}
 Hence the  exponential convergence of the sequence $F_n$
  in the $\|\cdot\|_{p,U}$ norm follows from \eqref{poorman.tm.rem2.eq4} and \eqref{poorman.tm.rem2.eq7}.

{\bf Numerical stability and stopping rule}:\quad
  Next let us consider the numerical stability of the  iterative algorithm \eqref{poorman.tm.rem2.eq3}. 
  Assume that the numerical error
  in $n$-th iterative step in the  iterative  algorithm \eqref{poorman.tm.rem2.eq3} is
  $\epsilon_n, n\ge 0$, i.e.,
  \begin{equation}\label{poorman.tm.rem2.eq9}
  \left\{\begin{array}{l}\tilde F_0=Ac_0+\epsilon_0\\
  \tilde F_n= \tilde F_0+ (I-A)\tilde F_{n-1}+\epsilon_n, \quad \ n\ge 1.
  \end{array}\right.\end{equation}
Let $F_n=(f_n(\gamma))_{\gamma\in \Gamma}, n\ge 0$, where $f_n, n\ge 0$, are given in the iterative approximation-projection  reconstruction algorithm \eqref{poorman.tm.eq2} with initial
data $c_0$.
By induction, we obtain
\begin{equation}\label{poorman.tm.rem2.eq11}
\tilde F_n-F_n=-\sum_{k=0}^{n-1} (I-A)^{n-1-k} A\tilde \epsilon_k+\tilde \epsilon_n,
\end{equation}
where $\tilde \epsilon_0=\epsilon_0$ and
 $\tilde \epsilon_k=(k+1) \epsilon_0+\epsilon_1+\cdots+\epsilon_k$ for $k\ge 1$.
 Therefore
 \begin{eqnarray}\label{poorman.tm.rem2.eq12}
 & & \|\tilde F_n-F_n\|_{p,U}\\
   & \le &  \sum_{k=0}^{n-1}\|(I-A)^{n-1-k} A\tilde \epsilon_k\|_{p,U}+
 \|\tilde \epsilon_n\|_{p,U}\nonumber\\
 & \le & \sum_{k=0}^{n-1}  C_0 r_0^{n-1-k}
 \|\tilde \epsilon_k\|_{p,U}+\|\tilde \epsilon_n\|_{p,U}\nonumber\\
 & \le &  C_0 \sum_{k=0}^{n-1}  r_0^{n-1-k}\Big( (k+1)
 \|\epsilon_0\|_{p,U}
 +  \sum_{j=1}^k\|\epsilon_j\|_{p,U}\Big)
  \nonumber\\
   & & \quad +
 (n+1)\|\epsilon_0\|_{p,U}+ \sum_{j=1}^n \|\epsilon_j\|_{p,U}\nonumber\\
 & \le & \frac{1-r_0+C_0}{1-r_0} \Big((n+1)\|\epsilon_0\|_{p,U}+\sum_{j=1}^n\|\epsilon_j\|_{p,U}\Big).\nonumber
  \end{eqnarray}
Denote the limit of $F_n$  as $n$ tends to infinity by $F_\infty$. By  \eqref{poorman.tm.rem2.eq4} and \eqref{poorman.tm.rem2.eq7}
we have
\begin{equation}\label{poorman.tm.rem2.eq10}
\|F_n-F_\infty\|_{p,U} \le  \sum_{k=n}^\infty   C_0 r_0^{k+1}
\big\|c_0\big\|_{p,U} \le   \frac{C_0 r_0 }{1-r_0} r_0^{n}\big\|c_0\big\|_{p,U}.
\end{equation}
Define the  sample-to-numerical-error ratio (SNER) of the iterative algorithm \eqref{poorman.tm.rem2.eq9}
in the logarithmic decibel scale by
\begin{equation}\label{poorman.tm.rem2.eq13}
{\rm SNER (dB)}=20\inf_{n\ge 1} \log_{10} \frac{n \|c_0\|_{p,U}}{ n\|\epsilon_0\|_{p,U}+ \sum_{j=1}^n\|\epsilon_j\|_{p,U}}.
\end{equation}
Then
\begin{equation} \label{poorman.tm.rem2.eq14}
\|\tilde F_n-F_n\|_{p,U}\le \frac{1-r_0+C_0}{1-r_0}  (n+1) 10^{-{\rm SNER (dB)}/20}\|c_0\|_{p,U},
\end{equation}
which together with
 \eqref{poorman.tm.rem2.eq10}
 implies
 that
 \begin{equation} \label{poorman.tm.rem2.eq15}
 \|\tilde F_n-F_\infty\|_{p,U}\le  \frac{1-r_0+C}{1-r_0} \Big( r_0^{n+1}+ (n+1) 10^{-{\rm SNER (dB)}/20}\Big)\|c_0\|_{p,U}.
 \end{equation}
This suggests that  a
 reasonable stopping step $n_1$ in the iterative algorithm \eqref{poorman.tm.rem2.eq3}
  is
  \begin{equation}\label{poorman.tm.rem2.eq16}
  n_1=  \Big[\frac{{\rm SNER} ({\rm dB})}{20\log_{10} (1/r_0)}-\frac{\log_{10} (\ln (1/r_0))}{\log_{10}{1/r_0}}-1\Big],
  \end{equation}
as  the function $f(y)= r_0^y+y10^{-{\rm SNER (dB)}/20}$
  attains the absolute minimum at
  \begin{equation}
  y_0:=
  \frac{{\rm SNER} ({\rm dB})}{20\log_{10} (1/r_0)}-\frac{\log_{10} (\ln (1/r_0))}{\log_{10}{1/r_0}}.\end{equation}
\end{rem}

\section{Iterative frame reconstruction algorithm}
\label{framealgorithm.section}

In this section, we study the convergence and consistency of the  iterative  frame algorithm for
reconstructing a signal in the reproducing kernel subspace of
$L^p(\Rd)$  associated with an idempotent integral operator
from its samples taken a relatively-separated set with
sufficient small gap.
The readers may refer to \cite{ctmj00, cbook03} for an introduction to frame theory, and \cite{agsiam01, aunfao94, bbook92,
 cisieee98, lwjfaa96, nwmcss91,  yic67}  for various frame algorithms to  reconstruct a signal from its samples.

\begin{thm}\label{frame.tm}
Let $1\le p\le \infty$, $T$ be an  idempotent integral operator whose kernel $K$  satisfies
 \eqref{kernel.assumption1} and \eqref{kernel.assumption2},
$V$ be the reproducing kernel subspace of $L^p(\Rd)$ associated with the operator $T$, and
$\delta_1>0$ be so chosen that
\begin{equation}\label{frame.tm.eq2}
r_2 :=
(2 r_1 +r_0)r_0<1,
\end{equation}
where
$$r_0:=\big\|\sup_{z\in \Rd} |\omega_{\delta_1/2}(K)(\cdot+z,z)|\big\|_{L^1(\Rd)}$$
and
$$
 r_1:=\Big\|\sup_{z\in \Rd} |K(\cdot+z,z)|\Big\|_{L^1(\Rd)}. $$
% +\big\|\sup_{z\in \Rd} |\omega_{\delta_0/2}(K) (\cdot+z, z)|\big\|_1\Big)\\
% & &\quad  \times \big\|\sup_{z\in \Rd} |\omega_{\delta_0/2}(K) (\cdot+z, z)|\big\|_1<1\nonumber \end{eqnarray}
 Let $\Gamma$ be a relatively-separated subset of $\Rd$ with  gap  $\delta_1$,
 $U=\{u_\gamma\}_{\gamma\in \Gamma}$ be a BUPU associated with  the covering $\{\gamma+[-\delta_1/2, \delta_1/2]^d\}_{\gamma\in \Gamma}$,
 and
 \begin{equation}\label{frame.tm.eq3}
 S_{\Gamma, U} f(x):=\sum_{\gamma\in \Gamma}
(Tf)(\gamma) \|u_\gamma\|_{L^1(\Rd)} K(x, \gamma),
% \int_{\Rd} \Big(\sum_{\gamma\in \Gamma} K(x, \gamma) \| u_\gamma\|_1 K(\gamma, y) \Big) f(y) dy,
 \quad \ f\in L^p(\Rd)
 \end{equation}
 be the preconditioned frame operator  on $L^p(\Rd)$.
Given a sequence $c_0=(c_0(\gamma))_{\gamma\in \Gamma}\in
\ell^p(\Gamma)$, we define the iterative frame reconstruction
algorithm by
 \begin{equation}\label{frame.tm.eq3b}\left\{\begin{array}{l}
 f_0=\sum_{\gamma\in \Gamma} c_0(\gamma) \|u_\gamma\|_{L^1(\Rd)} K(\cdot,
 \gamma),\\
 f_{n}= f_0+ f_{n-1}-S_{\Gamma, U}f_{n-1}, \quad   n\ge 1.
 \end{array}\right.\end{equation}
 Then the iterative algorithm \eqref{frame.tm.eq3b} converges to
 $f_\infty$ exponentially and is consistent. Moreover,
 \begin{equation}\label{frame.tm.eq3c} f_\infty=R_{\rm F} f_0,
 \end{equation}
 where
   \begin{equation} \label{frame.tm.eq4}
 R_{\rm F}:= T+\sum_{n=1}^\infty (T-S_{\Gamma,U})^n
 \end{equation}
defines a bounded integral operator on $L^p(\Rd)$ and
 is a pseudo-inverse of the preconditioned frame operator $S_{\Gamma,U}$, i.e.,
 \begin{equation}\label{frame.tm.eq5}
 R_{\rm F} T=T R_{\rm F}= R_{\rm F} \quad {\rm and} \quad
 R_{\rm F} S_{\Gamma,U}= S_{\Gamma,U} R_{\rm F}=T.
 \end{equation}
 Furthermore, the  kernel $K_F(x,y)$ of the integral operator
  $R_{\rm F}$ satisfies \eqref{kernel.assumption1}, \eqref{kernel.assumption2}, and
 \begin{equation}\label{frame.tm.eq6}
 K_F(x,y)=\int_{\Rd} \int_{\Rd} K(x,z_1) K_F(z_1, z_2) K(z_2, y) dz_1 dz_2 \quad {\rm for \ all} \ x,y\in \Rd.
 \end{equation}
\end{thm}

\begin{proof} %[Proof of Theorem \ref{frame.tm}]\
Define an integral operator $C_{\Gamma, U}$ by
\begin{eqnarray}\label{frame.tm.pf.eq2}\quad
C_{\Gamma, U} f(x) & = & \int_{\Rd}\int_{\Rd} \Big(\sum_{\gamma\in \Gamma} \big( K(x, \gamma)-K(x,z)\big) u_\gamma(z)\\
 & & \quad \times \big(K(\gamma,y)- K(z, y)\big)\Big) f(y) dy dz \quad {\rm for \ all}  \ f\in L^p,\nonumber
\end{eqnarray}
and let $Q_{\Gamma, U}^*$ be the adjoint of the integral operator $Q_{\Gamma, U}$ in \eqref{poorman.tm.pf1.eq1}, i.e.,
%
%
% Define integral operators $A_{\Gamma,U}, B_{\Gamma, U}$ and $C_{\Gamma, U}$ by
%\begin{equation}\label{frame.tm.pf.eq1}
%A_{\Gamma, U} f(x)=\int_{\Rd}\Big(\sum_{\gamma\in \Gamma}  u_\gamma(x) \big(K(\gamma,y)- K(x, y)\big)\Big) f(y) dy, \end{equation}
\begin{equation} \label{frame.tm.pf.eq3}
Q_{\Gamma, U}^* f(x)=\int_{\Rd}
\Big(\sum_{\gamma\in \Gamma} \big(K(\gamma, x)-K(y,x)\big)  u_\gamma(y)\Big)
 f(y) dy  \quad {\rm for \ all}  \ f\in L^p. \end{equation}
Then
\begin{equation} \label{frame.tm.pf.eq4} S_{\Gamma, U}-T= T Q_{\Gamma,U} + Q_{\Gamma,U}^* T +C_{\Gamma,U},\end{equation}
which implies that
\begin{eqnarray} \label{frame.tm.pf.eq5}
 & & \|S_{\Gamma, U}f-Tf\|_p \\
& \le &
 \|T\| \|Q_{\Gamma,U}f\|_p+ \|Q_{\Gamma,U}^* Tf\|_p+\|C_{\Gamma, U} f\|_p\nonumber\\
 & \le &   \|T\| \Big\|\int_{\Rd} h_{\delta_1/2}(\cdot-y) |f(y)| dy\Big\|_p\nonumber\\
    & & +\Big\|\int_{\Rd} h_{\delta_1/2}(z-\cdot) |Tf(z)| dz\Big\|_p\nonumber\\
& & +    \Big\|\int_{\Rd}\int_{\Rd} h_{\delta_1/2}(\cdot-z)h_{\delta_1/2}(z-y) |f(y)| dydz\Big\|_p\nonumber\\
  &\le  &  r_2 \|f\|_p \quad {\rm for \ all} \ f\in V,\nonumber
  %(\|h_{\delta_0/2}\|_1^2+2 \|T\|\|h_{\delta_0/2}\|_1)\|f\|_p
 \end{eqnarray}
where  $h_\delta=\sup_{z'\in \Rd}\omega_{\delta}(K)(\cdot+z',z')$.

By the iterative algorithm \eqref{frame.tm.eq3b},
\begin{equation}
f_{n}= f_0+\sum_{k=1}^n (T-S_{\Gamma, U})^k f_0 \quad \ {\rm for \
all} \ n\ge 1.
\end{equation}
This together with \eqref{frame.tm.pf.eq5} proves the exponential
convergence of $f_n, n\ge 0$, and the limit function $f_\infty$ is
given by \eqref{frame.tm.eq3c}.

\bigskip

By \eqref{t2=t}, \eqref{frame.tm.eq3} and Theorem \ref{kernel.tm}  in  the  Appendix, we have
 \begin{equation} \label{frame.tm.pf.eq6}
 S_{\Gamma,U}T=TS_{\Gamma, U}=S_{\Gamma, U}.
 \end{equation}
 This together with the exponential convergence of the right hand side of
 the equation \eqref{frame.tm.eq4} establishes that $R_{\rm F}$
 is a bounded operator and satisfies
  \eqref{frame.tm.eq5}, and hence  it
 is the
  pseudo-inverse of $S_{\Gamma,U}$.

  \bigskip

  The consistency of the frame iterative algorithm \eqref{frame.tm.eq3b} follows from \eqref{frame.tm.eq3c}
  and the fact that $f_0=S_{\Gamma, U} g$ if the initial data $c_0=(g(\gamma))_{\gamma\in \Gamma}$ is the sample of $g\in V$ taken on the set $\Gamma$.

\bigskip

From \eqref{kernel.assumption1}, \eqref{frame.tm.eq2}, \eqref{frame.tm.pf.eq2}, \eqref{frame.tm.pf.eq3} and \eqref{frame.tm.pf.eq4}, it follows that
\begin{eqnarray*} %\label{error.lem3.pf.eq1}
\big\|\sup_{z'\in \Rd} | K_F(\cdot+z',z')|\big\|_1
  \le   \big\|\sup_{z'\in \Rd} | K(\cdot+z',z')|\big\|_1 +\sum_{n=1}^\infty (r_2)^n
 <\infty.
% \Big(\big\|\sup_{z'\in \Rd} |\omega_{\delta_1/2}(K) (\cdot+z', z')|\big\|_1^2\Big)^n \nonumber\\
% & & \quad \times \Big(2\big\|\sup_{z'\in \Rd} |K(\cdot+z',z')|\big\|_1+
% \big\|\sup_{z'\in \Rd} |\omega_{\delta_1/2}(K) (\cdot+z', z')|\big\|_1\Big)^n \nonumber \\
\end{eqnarray*}
Hence $K_F$ satisfies the off-diagonal decay property \eqref{kernel.assumption1}.
The reproducing equality \eqref{frame.tm.eq6} follows from
$$TR_{\rm F} T=R_{\rm F}$$
by \eqref{frame.tm.eq5}.
The regularity property \eqref{kernel.assumption2} for the kernel $K_F$
 holds because of the off-diagonal decay property \eqref{kernel.assumption1} for the kernel $F$,
the regularity property \eqref{kernel.assumption2} for the kernel $K$ of the idempotent operator $T$, and the following estimate
\begin{eqnarray*}
\omega_\delta(K_F)(x,y)
 & \le &  \int_{\Rd} \int_{\Rd}
\omega_\delta(K)(x, z_1) |K_F(z_1, z_2)|\\
& & \quad \times \big(|K(z_2, y)| +\omega_\delta(K)(z_2,y)\big) dz_1dz_2
\\
 & & +
\int_{\Rd} \int_{\Rd}
|K(x, z_1)| |K_F(z_1, z_2)||\omega_\delta(K)(z_2, y)| dz_1dz_2
%\\
% & & + \int_{\Rd} \int_{\Rd}
%\omega_\delta(K)(x, z_1) |K_F(z_1, z_2)|\\
%& & \times \big(|K(z_2, y)| +\omega_\delta(K)(z_2,y)\big) dz_1dz_2
\end{eqnarray*}
by \eqref{frame.tm.eq6}.
  \end{proof}

\bigskip
\section{Asymptotic pointwise error estimates for reconstruction algorithms} % in the Presence of White Noise}

In this section, we discuss the asymptotic pointwise error estimate for
reconstructing a signal %in a reproducing kernel subspace of
%$L^p(\Rd)$
from its samples corrupted by white noises, as
the maximal gap of the sampling set tends to zero.

\begin{thm}\label{error.tm}
Let $1\le p\le \infty$, $T$ be an idempotent
 integral operator  whose kernel $K$  satisfies \eqref{kernel.assumption1}
and
\eqref{kernel.assumption2}, % and the symmetric condition \ref{
and $V$ be the reproducing kernel subspace of $L^p(\Rd)$ associated
with the operator $T$.
 Let $\Gamma$ be a relatively-separated subset of $\Rd$ with gap
 $\delta$,
 $U:=\{u_\gamma\}_{\gamma\in \Gamma}$ be a BUPU associated with  the covering $\{\gamma+[-\delta/2, \delta/2]^d\}_{\gamma\in
 \Gamma}$, and
$R:= \{R_\gamma(x)\}_{\gamma\in \Gamma}$  be either
  the displayer $\{(\|u_\gamma\|_{L^1(\Rd)})^{-1} R_{\rm AP} u_\gamma\}_{\gamma\in \Gamma}$ in the approximation-projection reconstruction
   algorithm
   or
 the displayer $\{R_{\rm F} K(\cdot, \gamma)\}_{\gamma\in \Gamma}$ in the frame  reconstruction algorithm where
  the operators $R_{\rm AP}$  and $R_{\rm F}$ are defined in \eqref{poorman.tm.pf1.eq9} and
  \eqref{frame.tm.eq4} respectively.
 Assume that
$\epsilon(\gamma), \gamma\in \Gamma$,  are bounded
i.i.d. noises with zero
mean and $\sigma^2$ variance, i.e.,
\begin{equation}\label{error.tm.eq3}
\epsilon(\gamma)\in [-B, B],\  E(\epsilon(\gamma))=0,\ \ {\rm  and}\   \
{\rm Var}(\epsilon(\gamma))=\sigma^2\end{equation} for some
positive constant $B$, and that the initial data $c_0$ is the
sample of a signal $g\in V$  taken on $\Gamma$ corrupted by
random noise $\epsilon:=(\epsilon(\gamma))_{\gamma\in \Gamma}$, i.e.,
\begin{equation} \label{error.tm.eq4} c_0= (g(\gamma)+\epsilon(\gamma))_{\gamma\in \Gamma}.\end{equation}
 Then for any $x\in \Rd$
 \begin{equation}\label{error.tm.eq5}
 E\big(g(x)-  Rc_0 (x)\big)=0
 %\sum_{\gamma\in \Gamma} (g(\gamma)+\epsilon(\gamma))\|u_\gamma\|_1 (S_{\Gamma,U}^\dag K(\cdot,\gamma))(x)\Big)=0,
 \end{equation}
 and
  \begin{eqnarray}\label{error.tm.eq6}
 & &  {\rm Var} \big(g(x)- R c_0(x)\big)=\sum_{\gamma\in \Gamma} \|u_\gamma\|_{L^1(\Rd)}^2 |R_\gamma(x)|^2
 %\sum_{\gamma\in \Gamma} (g(\gamma)+\epsilon(\gamma)) \|u_\gamma\|_1 (S_{\Gamma,U}^\dag K(\cdot,\gamma))(x)\Big)\\
  \\
  & \le  &  \sigma^2 \sup_{\gamma\in \Gamma} \|u_\gamma\|_{L^1(\Rd)}  \Big(\int_{\Rd}
|K(x,z)|^2 dz+ o(1)\Big) \quad \ {\rm as}\ \delta\to 0,\nonumber
 \end{eqnarray}
where
\begin{equation}
\label{error.tm.eq6b}
Rc_0 (x) =\sum_{\gamma\in \Gamma} c_0(\gamma) \|u_\gamma\|_{L^1(\Rd)} R_\gamma(x)\quad {\rm for \ all} \  c_0=(c_0(\gamma))_{\gamma\in \Gamma}\in \ell^\infty(\Gamma).
\end{equation}
Furthermore if
 \begin{equation}\label{error.tm.eq7} \|u_{\gamma}\|_{L^1(\Rd)}=\alpha(\delta)  (1+o(1)) \quad {\rm as} \ \delta\to 0\end{equation}
 for some  positive numbers $\alpha(\delta)$ independent of $\gamma$, then
 the inequality in \eqref{error.tm.eq6} becomes an equality, i.e.,
 \begin{eqnarray}
\label{error.tm.eq8}
  {\rm Var} \big(g(x)- R c_0(x)\big)
  %\sum_{\gamma\in \Gamma} (g(\gamma)+\epsilon(\gamma))\|u_\gamma\|_1 (S_{\Gamma,U}^\dag K(\cdot,\gamma))(x)\Big)\\
  =  \alpha(\delta) \sigma^2
 \Big(\int_{\Rd} |K(x,z)|^2 dz +o(1)\Big)
 \end{eqnarray}
  as $\delta$ tends to zero.
\end{thm}

\begin{rem} \label{error.rem1} The error estimate \eqref{error.tm.eq8}  is established in \cite{alsieee08} for
reconstructing signals  in  a finitely-generated
 shift-invariant subspace of $L^2(\Rd)$ from   corrupted uniform sampling data via  the frame reconstruction algorithm. More precisely,
 $\Gamma=\delta\Zd$,
 $u_\gamma(x)=\chi_{[-\delta/2, \delta/2]^d} (x-\gamma)$ for $\gamma\in \Gamma$,
 the  idempotent operator $T$ is defined in \eqref{sisoperator.def},
%$$Tf(x)=\int_{\Rd} \Big(\sum_{i=1}^r \sum_{k\in \Zd}
%\phi_i(x-k)\phi_i(y-k)\Big) f(y) dy,\quad  f\in L^2(\Rd)$$
and the range space associated with the idempotent operator $T$ is the shift-invariant space $V_2(\phi_1, \ldots, \phi_r)$
in \eqref{sis.def}.
\end{rem}

\begin{rem}\label{error.rem2}
 By the definition of a BUPU  associated with the covering $\{\gamma+[-\delta/2, \delta/2]^d\}_{\gamma\in \Gamma}$ of $\Rd$, we have
\begin{equation}\label{error.rem2.eq1}\|u_\gamma\|_{L^1(\Rd)}\le \delta^d.\end{equation}
The above inequality becomes an equality when $\Gamma=\delta\Zd$ and $u_\gamma=\chi_{[-\delta/2, \delta/2]^d}$.
It is expensive  to find the operators $R_{\rm AP}$ and $R_{\rm F}$
when the sampling set has very small gap $\delta$. As noticed in the proof of Theorem \ref{error.tm}, both operators
are  close to the idempotent operator $T$ when the sampling set has very small gap.
Then a natural replacement of  the displayer $R_\gamma$ in \eqref{error.tm.eq6b}
is either $(\|u_\gamma\|_{L^1(\Rd)})^{-1} Tu_\gamma$ or $K(\cdot, \gamma)$.
In both cases, the variance estimates in \eqref{error.tm.eq6} and \eqref{error.tm.eq8}
still hold, but the unbiased condition \eqref{error.tm.eq6} does not.
 \end{rem}

To prove Theorem \ref{error.tm}, we need several technical lemmas.
The first lemma is a slight generalization of Theorem \ref{error.tm}.

\begin{lem}\label{error.lem1}
Let  the operator $T$, the kernel $K$, the reproducing kernel space $V$, the sampling set $\Gamma$,  the bounded uniform partition of unity $U=\{u_\gamma\}_{\gamma\in \Gamma}$,
the random noise $\epsilon$, and the variance $\sigma$ of the noise $\epsilon$ be as in Theorem \ref{error.tm}, and let the displayer
$R:= \{R_\gamma(x)\}_{\gamma\in \Gamma}$ satisfy
\begin{equation}\label{error.lem1.eq1}
g(x)= \sum_{\gamma\in \Gamma} g(\gamma) \|u_\gamma\|_{L^1(\Rd)} R_\gamma(x) \quad {\rm for \ all} \ g\in V,
\end{equation}
and
\begin{equation}\label{error.lem1.eq2}
\lim_{\delta\to 0}
\big\| \sup_{\gamma\in \Gamma} \sup_{z\in \gamma+[-\delta/2, \delta/2]^d}
|R_\gamma (\cdot+z)-K(\cdot+z, z)|\big\|_{L^1(\Rd)}=0.
\end{equation}
Then \eqref{error.tm.eq5}, \eqref{error.tm.eq6} and \eqref{error.tm.eq8} hold.
\end{lem}

\begin{proof}
Set
\begin{equation}\label{error.lem1.pf.eq0}
h_\delta(x)=\sup_{\gamma\in \Gamma} \sup_{z\in \gamma+[-\delta/2, \delta/2]^d}
|R_\gamma (x+z)-K(x+z, z)|.\end{equation}
By \eqref{kernel.assumption1}, \eqref{error.lem1.eq2} and \eqref{error.lem1.pf.eq0},   we have
\begin{eqnarray}\label{error.lem1.pf.eq1}
% &  &
 \sum_{\gamma\in \Gamma} \|u_\gamma\|_1 |R_\gamma(x)| % &\\
 & \le & \int_{\Rd} \sum_{\gamma\in \Gamma} u_\gamma(z)\big (|K(x,z)|+h_\delta(x-z)\big) dz\\ %\nonumber\\
  & \le &
 \big\|\sup_{z\in \Rd} |K(\cdot+z, z)|\big\|_1+\|h_\delta\|_1<\infty.\nonumber
 \end{eqnarray}
 This together with \eqref{error.tm.eq3} and \eqref{error.lem1.eq1}  leads to
 \begin{eqnarray} \label{error.lem1.pf.eq2}
 E\big(g(x)-  Rc_0 (x)\big) & = &
 E\Big(\sum_{\gamma\in \Gamma} \epsilon(\gamma) \|u_\gamma\|_1 R_\gamma(x)\Big)\\
  & = &
 \sum_{\gamma\in \Gamma} E(\epsilon(\gamma)) \|u_\gamma\|_1 R_\gamma(x)=0,
 \nonumber \end{eqnarray}
and the unbiased property
\eqref{error.tm.eq5}  for the reconstruction process in \eqref{error.tm.eq6b} follows.

\bigskip

By \eqref{error.tm.eq3},  \eqref{error.tm.eq5} and \eqref{error.lem1.pf.eq1}, we obtain
\begin{eqnarray}\label{error.lem1.pf.eq3}
 {\rm Var} \big(g(x)-  Rc_0 (x)\big)
 &= &  E\Big(\sum_{\gamma\in \Gamma} \epsilon(\gamma) \|u_\gamma\|_1 R_\gamma(x)\Big)^2\nonumber\\
 & = & \sigma^2 \sum_{\gamma\in \Gamma}  \|u_\gamma\|_1^2 |R_\gamma(x)|^2.\nonumber
\end{eqnarray}
Therefore
\begin{eqnarray} \label{error.lem1.pf.eq4}
   & & {\rm Var} \big(g(x)-  Rc_0 (x)\big)\\
  & \le  & \sigma^2\big( \sup_{\gamma\in \Gamma} \|u_\gamma\|_1\big)
 \Big(\sum_{\gamma\in \Gamma} \|u_\gamma\|_1 |R_\gamma(x)|^2\Big)\nonumber\\
 & \le  & \sigma^2\big( \sup_{\gamma\in \Gamma} \|u_\gamma\|_1\big)
  \Big(\int_{\Rd} \big(|K(x,z)|+ |h_\delta(x-z)|\big)^2 dz\Big)\nonumber\\
 & \le &
  \sigma^2\big( \sup_{\gamma\in \Gamma} \|u_\gamma\|_1\big)
  \Big(\int_{\Rd}|K(x,z)|^2 dz+ o(1)\Big),\nonumber
 \end{eqnarray}
 where we have used  \eqref{error.lem1.eq2}  and \eqref{error.lem1.pf.eq0} to obtain the last two estimates.
 Hence the variance estimate \eqref{error.tm.eq6}  for the reconstruction process in \eqref{error.tm.eq6b} is established.

\bigskip

By \eqref{error.tm.eq7}, \eqref{error.lem1.eq2} and \eqref{error.lem1.pf.eq3}, we get
\begin{eqnarray}
 & & {\rm Var} \big(g(x)-  Rc_0 (x)\big)\\
 & = &   \sigma^2 \big(\alpha(\delta)+o(1)\big)  \Big(\sum_{\gamma\in \Gamma} \|u_\gamma\|_1 |R_\gamma(x)|^2\Big)\nonumber\\
 & = & \sigma^2 \big(\alpha(\delta)+o(1)\big)  \Big(\int_{\Rd} \big(K(x,z)+O(h_\delta(x-z))\big)^2  dz\Big)\nonumber\\
 & = & \sigma^2 \alpha(\delta)\Big (\int_{\Rd} |K(x,z)|^2 dz+o(1)\Big),\nonumber
 \end{eqnarray}
 and hence \eqref{error.tm.eq8} is proved.
\end{proof}

\begin{lem}\label{error.lem2}
 Let  the operator $T$, the kernel $K$, the reproducing kernel space $V$, the sampling set $\Gamma$,  the bounded uniform partition of unity $U=\{u_\gamma\}_{\gamma\in \Gamma}$,
the random noise $\epsilon$, and the variance $\sigma$ of the noise $\epsilon$ be as in Theorem \ref{error.tm},
and let the displayer
$R=\{R_\gamma\}_{\gamma\in \Gamma}$ be defined by
\begin{equation}\label{error.lem2.eq1}
R_\gamma= (\|u_\gamma\|_1)^{-1} R_{\rm AP} u_\gamma, \gamma\in \Gamma\end{equation}
where $R_{\rm AP}$ is given in \eqref{poorman.tm.pf1.eq9}.
Then  the above displayer $R$ satisfies \eqref{error.lem1.eq1} and \eqref{error.lem1.eq2}.
\end{lem}

\begin{proof}
By
\eqref{poorman.tm.pf1.eq9}, \eqref{poorman.tm.pf1.eq12} and \eqref{poorman.tm.pf1.eq13},
the reconstruction formula \eqref{error.lem1.eq1}
holds for  the displayer $R$ in \eqref{error.lem2.eq1}.

Denote the kernel of the integral operators $R_{\rm AP}-T$ by $\tilde K_{\rm AP}$.
By \eqref{t2=t},   \eqref{poorman.tm.pf1.eq3b}, \eqref{poorman.tm.pf1.eq5b}, \eqref{poorman.tm.pf1.eq9} and \eqref{apkernel.rem.eq1}, we have
\begin{equation}\label{error.lem2.pf.eq1}
\tilde K_{AP}(x,y)=\int_{\Rd} \int_{\Rd} K(x,z_1) \tilde K_{\rm AP}(z_1, z_2) K(z_2, y) dz_1dz_2,
\end{equation}
and
\begin{eqnarray} \label{error.lem2.pf.eq2} &  &   \big\|\sup_{z'\in \Rd} |\tilde K_{\rm AP}(\cdot+z', z')|\big\|_1\\
& \le & \sum_{n=1}^\infty \big\| \sup_{z'\in \Rd} |K(\cdot+z',z')|\big\|_1 \Big(
\big\|\sup_{z'\in \Rd} |\omega_{\delta/2}(K) (\cdot+z',z')|\big\|_1\Big)^n\nonumber\\
& \to & 0 \quad {\rm as} \ \delta \to 0. \nonumber\end{eqnarray}
This together with \eqref{kernel.assumption1} and \eqref{kernel.assumption2}
 implies that
\begin{eqnarray}\label{error.lem2.pf.eq3}
 & &\Big\| \sup_{\gamma\in \Gamma}
 \sup_{z'\in \gamma+[-\delta/2, \delta/2]^d}\big| (\|u_\gamma\|_1)^{-1} R_{\rm AP} u_\gamma (\cdot+z')-K(\cdot+z',z')\big |\Big\|_1\\
 & \le &\Big\|\sup_{z'\in \Rd}
 \omega_{\delta}(K)(\cdot+z',z')\Big\|_1+\Big\|\sup_{z\in \Rd} \int_{\Rd}\int_{\Rd} |K(\cdot+z,z_1)| \nonumber\\
  & & \quad \times |\tilde K_{\rm AP}(z_1, z_2)|\big (|K(z_2, z)|+|\omega_{\delta}(K)(z_2, z)|\big)dz_1 dz_2\Big\|_1\nonumber\\
   & \to & 0\quad {\rm as} \ \delta\to 0.\nonumber
\end{eqnarray}
Hence  \eqref{error.lem1.eq2} follows.
\end{proof}

\begin{lem}\label{error.lem3}
 Let  the operator $T$, the kernel $K$, the reproducing kernel space $V$, the sampling set $\Gamma$,  the bounded uniform partition of unity $U=\{u_\gamma\}_{\gamma\in \Gamma}$,
the random noise $\epsilon$, and the variance $\sigma$ of the noise $\epsilon$ be as in Theorem \ref{error.tm},
and let the displayer
$R=\{R_\gamma\}_{\gamma\in \Gamma}$ be defined by
\begin{equation}\label{error.lem3.eq1}
R_\gamma= R_{\rm F} K(\cdot, \gamma),\quad  \gamma\in \Gamma\end{equation}
where $R_{\rm F}$ is given in \eqref{frame.tm.eq4}.
Then the above displayer $R$ satisfies \eqref{error.lem1.eq1} and \eqref{error.lem1.eq2}.
\end{lem}

\begin{proof}
The reconstruction formula \eqref{error.lem1.eq1} follows from  Theorem \ref{frame.tm}.

Denote the integral kernel of the integral operator $R_{\rm F}-T$ by
$\tilde K_F$.
Then
\begin{equation} \tilde K_F(x,y)=\int_{\Rd} \int_{\Rd} K(x,z_1) \tilde K_F(z_1, z_2) K(z_2, y) dz_1 dz_2, \end{equation}
and
\begin{eqnarray}\label{error.lem3.pf.eq1}
 & & \big\|\sup_{z\in \Rd} |\tilde K_F(\cdot+z,z)|\big\|_1\\
 & \le &  \sum_{n=1}^\infty \Big(2\big\|\sup_{z\in \Rd} |K(\cdot+z,z)|\big\|_1+ \big\|\sup_{z\in \Rd} |\omega_{\delta}(K) (\cdot+z, z)|\big\|_1\Big)^n \nonumber \\
 & & \quad \times \Big(\big\|\sup_{z\in \Rd} |\omega_{\delta}(K) (\cdot+z, z)|\big\|_1^2\Big)^n \to 0 \quad \ {\rm as} \ \delta\to 0\nonumber
\end{eqnarray}
by \eqref{kernel.assumption2}, \eqref{frame.tm.eq4},  and \eqref{frame.tm.pf.eq4}.
Therefore
\begin{eqnarray*}
 & & % \limsup_{\delta\to 0}
\int_{\Rd}  \sup_{\gamma\in \Gamma} \sup_{z\in \gamma+[-\delta/2, \delta/2]^d}
|R_F K(\cdot,\gamma)) (x+z)-K(x+z, z)| dx\\
 & \le &
 %\limsup_{\delta\to 0}
\int_{\Rd}  \sup_{\gamma\in \Gamma} \sup_{z\in \gamma+[-\delta/2, \delta/2]^d}
|K(x+z,\gamma)  -K(x+z, z)| dx\\
 & & + % \limsup_{\delta\to 0}
\int_{\Rd}  \sup_{\gamma\in \Gamma} \sup_{z\in \gamma+[-\delta/2, \delta/2]^d}\\
& & \qquad
\Big|\int_{\Rd}\int_{\Rd} K(x+z,z_1) \tilde K_F(z_1, z_2) K(z_2, \gamma) dz_1 d_2\Big| dx\\
 & \le &
 %\limsup_{\delta\to 0}
\int_{\Rd}
\sup_{z'\in\Rd} |\omega_{\delta/2}(K)(x+z',z')| dx\\
 & & + %\limsup_{\delta\to 0}
\int_{\Rd}
\int_{\Rd}\int_{\Rd} \big(\sup_{z'\in \Rd} |K(x-z_1+z', z')|\big)
%\\
%& & \qquad\times
\big(\sup_{z'\in \Rd} |\tilde K_F(z_1- z_2+z', z')|\big) \\
& & \qquad \times \big(\sup_{z'\in \Rd}
|K(z_2+z', z')|+\sup_{z'\in \Rd}\omega_{\delta/2}(K)(z_2+z', z')| \big) dz_1 d_2 dx\\
& \to &  0\quad {\rm as} \ \delta\to 0.
\end{eqnarray*}
Then \eqref{error.lem1.eq2}
is established for the displayer $R$ in \eqref{error.lem3.eq1}.
 \end{proof}

%We start to prove Theorem \ref{error.tm}.

 \begin{proof} [Proof of Theorem \ref{error.tm}]
The conclusions in  Theorem \ref{error.tm} follows directly from Lemmas \ref{error.lem1}, \ref{error.lem2} and \ref{error.lem3}.
\end{proof}

\bigskip
\begin{appendix}
\section{Reproducing kernel subspaces of $L^p(\Rd)$ associated with idempotent integral operators} % of $L^p(\Rd)$}

The range space associated with an idempotent operator $T$ on $L^p(\Rd)$
whose kernel satisfies \eqref{kernel.assumption1} and \eqref{kernel.assumption2}
include the space of all $p$-integrable non-uniform splines of order $n$ satisfying $n-1$ continuity conditions at each knot (Example \ref{modelspace1.example}), and
 the space introduced in \cite{sunaicm08} for modeling signals with finite rate of innovation
(Example \ref{modelspace2.example}).
In this appendix, we establish some properties of  such range spaces, particularly
the reproducing kernel property in Theorem \ref{kernel.tm} and the frame property in Theorem \ref{localizedframeforrange.tm}.

\subsection{Reproducing kernel property}

In this subsection, we show that the range space of an idempotent operator on $L^p(\Rd)$ whose kernel satisfies \eqref{kernel.assumption1} and
 \eqref{kernel.assumption2} has
 some properties similar to  the ones for a reproducing kernel Hilbert
subspace of $L^2(\Rd)$.

\begin{thm}
\label{kernel.tm}
Let $T$ be an idempotent  integral operator on $L^p(\Rd)$  whose kernel $K$
 satisfies \eqref{kernel.assumption1} and \eqref{kernel.assumption2},
 and $V$ be the range space of the operator $T$.
 Set
 \begin{eqnarray*} a_\delta(q) & = & \delta^{-d+d/q} \big(\big\|\sup_{z\in \Rd} |K(\cdot+z,z)|\big\|_{L^1(\Rd)}\big)^{1/q}\nonumber\\
 & &  \times \Big(\big\|\sup_{z\in \Rd} |K(\cdot+z,z)|\big\|_{L^1(\Rd)}+\big\|\sup_{z\in \Rd} |\omega_\delta(K)(\cdot+z,z)|\big\|_{L^1(\Rd)}\Big)^{1-1/q}
 \end{eqnarray*}
 and
$$b_\delta(q)=  (6^d+1)^{1-1/q}\delta^{-d+d/q}\big\|\sup_{z\in \Rd} |\omega_\delta(K)(\cdot+z,z)|\big\|_{L^1(\Rd)}$$
for $\delta>0$ and $1\le q\le \infty$.
Then
\begin{itemize}

\item [{(i)}]
$V$ is a reproducing kernel subspace of $L^p(\Rd)$. Moreover,
\begin{eqnarray}\label{kernel.tm.eq1}
|f(x)| & \le &  a_\delta(p/(p-1))
\|f\|_{L^p(\Rd)}\nonumber
\end{eqnarray}
for any $f\in V$ and $\delta>0$.

\item[{(ii)}] The kernel $K$ satisfies the ``reproducing kernel property":
\begin{equation}\label{kernel.tm.eq2}\int_{\Rd} K(x,z) K(z,y) dz= K(x,y)\quad \ {\rm for \ all} \ x, y\in \Rd.\end{equation}

\item [{(iii)}] $K(\cdot,y)\in V$ for any $y\in \Rd$.

\item [{(iv)}]  The functions
$K(x, \cdot),\  K(\cdot, y),\  \omega_\delta(K)(x, \cdot)$ and $\omega_\delta(K)(\cdot, y)$ belong to $L^q(\Rd)$ for all $x, y\in \Rd$ and $1\le q\le \infty$, and their $L^q(\Rd)$-norms are uniformly bounded. Moreover,
\begin{eqnarray}\label{kernel.tm.eq3}
 & & \max\Big( \sup_{x\in \Rd}\|K(x, \cdot)\|_{L^q(\Rd)} , \ \sup_{y\in \Rd} \|K( \cdot, y)\|_{L^q(\Rd)}\Big)\le a_\delta(q)
 \end{eqnarray}
 and
 \begin{eqnarray}\label{kernel.tm.eq4}
 & &  \max \Big(\sup_{x\in \Rd} \|\omega_\delta(K)(x, \cdot)\|_{L^q(\Rd)}, \  \sup_{y\in \Rd}  \|\omega_\delta(K)( \cdot, y)\|_{L^q(\Rd)}\Big)\le b_\delta(q).\end{eqnarray}

\end{itemize}
\end{thm}

\begin{proof} %[Proof of Theorem \ref{kernel.tm}]
(iv):\
 Note that
 \begin{equation}\label{kernel.tm.pf.eq2}
 |K(x,y)|\le |K(x,z)|+|\omega_{\delta}(K)(x,z)|
 \end{equation}
 holds for  all $y, z\in k\delta+[-\delta/2, \delta/2]^d$  and $x\in \Rd$, where $k\in \Zd$ and  $\delta>0$, we then have
\begin{equation} \label{kernel.tm.pf.eq3}
|K(x,y)|\le \delta^{-d}  \int_{k\delta+[-\delta/2, \delta/2]^d}  \big(|K(x,z)|+|\omega_{\delta}(K)(x,z)|\big) dz.
\end{equation}
Thus
\begin{eqnarray*}
 & & \sup_{x\in \Rd} \|K(x, \cdot)\|_\infty  \\
& \le & \delta^{-d}  \sup_{x\in \Rd} \sup_{k\in \Zd} \int_{k\delta+[-\delta/2, \delta/2]^d}  \big(|K(x,z)|+|\omega_{\delta}(K)(x,z)|\big) dz\nonumber\\
&\le & \delta^{-d}\Big(\big\|\sup_{z\in \Rd} |K(\cdot+z,z)|\big\|_1+\big\|\sup_{z\in \Rd} |\omega_\delta(K)(\cdot+z,z)|\big\|_1\Big)\nonumber
\end{eqnarray*}
and
\begin{equation*} \label{kernel.tm.pf.eq5}
 \sup_{x\in \Rd}  \|K(x, \cdot)\|_1\le
\big\|\sup_{z\in \Rd} |K(\cdot+z,z)|\big\|_1.\nonumber
\end{equation*}
Interpolating the above  estimates for the $L^1$ and $L^\infty$ norms of $K(x, \cdot)$, we obtain
\begin{eqnarray}\label{kernel.tm.pf.eq4}
& & \sup_{x\in \Rd}  \|K(x, \cdot)\|_q  \le  a_\delta(q).
% \delta^{-d+d/q} \big(\big\|\sup_{z\in \Rd} |K(\cdot+z,z)|\big\|_1\big)^{1/q}\\
% & & \qquad \quad \times \Big(\big\|\sup_{z\in \Rd} |K(\cdot+z,z)|\big\|_1+\big\|\sup_{z\in \Rd} |\omega_\delta(K)(\cdot+z,z)|\big\|_1\Big)^{1-1/q}\nonumber
\end{eqnarray}
Similarly, we have
\begin{eqnarray} \label{kernel.tm.pf.eq6}
  & & \sup_{y\in \Rd} \|K(\cdot,y)\|_q \le  a_\delta(q).
%  \delta^{-d+d/q} \big(\big\|\sup_{z\in \Rd} |K(\cdot+z,z)|\big\|_1\big)^{1/q}\\
% & & \qquad \quad \times \Big(\big\|\sup_{z\in \Rd} |K(\cdot+z,z)|\big\|_1+\big\|\sup_{z\in \Rd} |\omega_\delta(K)(\cdot+z,z)|\big\|_1\Big)^{1-1/q}\nonumber
\end{eqnarray}
 Therefore \eqref{kernel.tm.eq3} follows.

 From the definition of  modulus of continuity, we obtain
\begin{eqnarray}\label{kernel.tm.pf.eq7}
 & & \sup_{y\in k\delta+[-\delta/2, \delta/2]^d}
\omega_\delta(K)(x,y) \\
& \le &  \inf_{z\in k\delta+[-\delta/2, \delta/2]^d} \omega_\delta(K) (x,z)+ \omega_{2\delta}(K)(x,z)\nonumber
\\
& \le & \delta^{-d} \int_{k\delta+[-\delta/2, \delta/2]^d} \big( \omega_\delta(K) (x,z)+
 \omega_{2\delta}(K)(x,z) \big) dz\nonumber\end{eqnarray}
 for any $x\in \Rd$ and $k\in \Zd$, and
  \begin{equation}\label{kernel.tm.pf.eq9}
\omega_{2\delta}(K)(x,y)\le \sum_{\epsilon, \epsilon'\in \{-1, 0, 1\}^d}
\omega_\delta (K) (x+\epsilon \delta, y+\epsilon'\delta)
%\big\|_1\le 3^d \big\|\sup_{z\in \Rd} |\omega_\delta(K)(\cdot+z,z)|\big\|_1.
\end{equation}
for all $x,y\in \Rd$.
By an argument similar to the one used in establishing \eqref{kernel.tm.pf.eq4} and \eqref{kernel.tm.pf.eq6} except we now use
\eqref{kernel.tm.pf.eq7} instead of \eqref{kernel.tm.pf.eq2} and apply \eqref{kernel.tm.pf.eq9}
to estimate $\|\sup_{z\in \Rd} \omega_{2\delta}(K)(\cdot+z,z)\|_1$, we then obtain
\begin{equation} \label{kernel.tm.pf.eq8*}%\quad
\sup_{x\in\Rd} \|\omega_\delta(K)(x, \cdot)\|_q
 %& \le &
%\delta^{-d+d/p}\Big(\big\|\sup_{z\in \Rd} |\omega_\delta(K)(\cdot+z,z)|\big\|_1+\big\|\sup_{z\in \Rd} |\omega_{2\delta}(K)(\cdot+z,z)|\big\|_1\Big)\\
% &\le &   \delta^{-d}
% \Big( \big\|\sup_{z\in \Rd} \omega_\delta(K)(\cdot+z,z)\big\|_1 +\big\|\sup_{z\in \Rd} \omega_{2\delta}(K)(\cdot+z,z)\big\|_1 \Big)\\
  \le  (6^d+1)^{1-1//q}  \delta^{-d+d/q}\big\|\sup_{z\in \Rd} \omega_\delta(K)(\cdot+z,z)\big\|_1  ,
\end{equation}
and
\begin{equation} \label{kernel.tm.pf.eq10}\quad
\sup_{y\in \Rd} \|\omega_\delta(K)(\cdot,y)\|_q %& \le &
%\delta^{-d+d/p}\Big(\big\|\sup_{z\in \Rd} |\omega_\delta(K)(\cdot+z,z)|\big\|_1+\big\|\sup_{z\in \Rd} |\omega_{2\delta}(K)(\cdot+z,z)|\big\|_1\Big)\\
 \le  (6^d+1)^{1-1//q}  \delta^{-d+d/q}\big\|\sup_{z\in \Rd} |\omega_\delta(K)(\cdot+z,z)|\big\|_1.
\end{equation}
Combining \eqref{kernel.tm.pf.eq8*} and \eqref{kernel.tm.pf.eq10} proves
 \eqref{kernel.tm.eq4}.

 \bigskip

 (i):\  By \eqref{integraloperator.def} and
\eqref{kernel.tm.eq3}, we have
\begin{equation*} |f(x)|  \le   \|K(x, \cdot)\|_{p/(p-1)} \|f\|_p\le   a_\delta(p/(p-1))\|f\|_p\end{equation*}
for all $x\in \Rd$ and $f\in V$. Then \eqref{kernel.tm.eq1} holds and
 $V$ is a reproducing kernel subspace of $L^p$.

 \bigskip

 (ii): Noting that
\begin{eqnarray*}
 \int_{\Rd} \sup_{z\in \Rd} \Big| \int_{\Rd} K(x+z,y) K(y,z) dy\Big| dx
 \le \Big(\int_{\Rd}\big(\sup_{z\in \Rd} |K(x+z,z)|\big) dx\Big)^2 <\infty,\nonumber
\end{eqnarray*}
we then have that the kernel
$$A(x,y):=\int_{\Rd} K(x+z,y) K(y,z) dy-K(x,y)$$
of the linear operator $T^2-T$ satisfies
$\big\|\sup_{z\in \Rd} |A(\cdot+z,z)|\big\|_1<\infty.$
This together with \eqref{t2=t} proves  \eqref{kernel.tm.eq2}.

\bigskip

 (iii): \ The conclusion that $K(\cdot, y)\in V$ for any $y\in \Rd$ follows from  \eqref{kernel.tm.eq2} and
  \eqref{kernel.tm.eq3}.
  \end{proof}

\subsection{Frame property}
In this subsection, we show that the range space of an idempotent
integral operator  whose kernel satisfies
\eqref{kernel.assumption1} and \eqref{kernel.assumption2} has localized frames.
Let $1\le p\le \infty$, $V\subset L^p$ and $W\subset L^{p/(p-1)}$. We say that
the $p$-frame   $\tilde \Phi=\{\tilde \phi_\lambda\}_{\lambda\in \Lambda}\subset W $ for
$V$ and
the $p/(p-1)$-frame   $\Phi=\{\phi_\lambda\}_{\lambda\in \Lambda}\subset V $ for $W$ form a {\em dual pair}
if the following reconstruction formulae hold:
\begin{equation}
f=\sum_{\lambda\in \Lambda} \langle f, \tilde \phi_\lambda\rangle \phi_\lambda \quad \ {\rm for \ all} \ f\in V,
\end{equation}
and
\begin{equation}
g=\sum_{\lambda\in \Lambda}\langle g, \phi_\lambda\rangle  \tilde \phi_\lambda \quad \ {\rm for\  all} \ g\in W.
\end{equation}
Here we denote by $\langle f, g\rangle$  the standard action \eqref{lplqaction.def} between a function $f\in L^p$ and a function $g\in L^{p/(p-1)}$.

 \begin{thm}\label{localizedframeforrange.tm}
Let $1\le p\le \infty$,  $T$ be an idempotent integral operator on $L^p(\Rd)$ whose
 kernel $K$  satisfies \eqref{kernel.assumption1} and \eqref{kernel.assumption2},
$T^*$ be the  adjoint of the idempotent operator $T$, i.e.,
\begin{equation}
T^* g(x)=\int_{\Rd} K(y,x) g(y) dy \quad \ {\rm for \ all} \ g\in L^{p/(p-1)}(\Rd),
\end{equation}
and let  $V$  and $V^*$ be the  range spaces of the operator $T$ on $L^p(\Rd)$ and the operator $T^*$ on $L^{p/(p-1)}(\Rd)$ respectively.
Then  there exist a relatively-separated
subset $\Lambda$, and two families $\Phi:=\{\phi_\lambda\}_{\lambda\in \Lambda}$ of functions $\phi_\lambda\in V$
and
$\tilde \Phi:=\{\tilde \phi_\lambda\}_{\lambda\in \Lambda}$ of functions $\tilde \phi_\lambda\in V^*$
such that
\begin{itemize}
\item[{(i)}] Both $\Phi$ and $\tilde \Phi$ are localized in the sense that
\begin{equation}\label{localizedframeforrange.tm.eq2}
\left\{\begin{array}
{l} |\phi_\lambda(x)|+|\tilde \phi_\lambda(x)|\le h(x-\lambda)\\
 |\omega_\delta(\phi_\lambda)(x)+\omega_\delta(\tilde \phi_\lambda)(x)|\le h_\delta (x-\lambda)
 \quad \ {\rm for\ all} \ \lambda\in \Lambda \ {\rm and} \ x\in \Rd.\end{array}\right.
\end{equation}
where $h$ and $h_\delta$ are integrable functions with
\begin{equation}\label{localizedframeforrange.tm.eq3}
\lim_{\delta\to 0} \|h_\delta\|_1=0,
\end{equation}

\item[{(ii)}] $\tilde \Phi$ is a $p$-frame for $V$ and $\Phi$ is a $p/(p-1)$-frame for $V^*$.

\item[{(iii)}]  $\Phi$ and $\tilde \Phi$ form a dual pair. %, i.e.,
% \begin{equation}
%  f(x)=\sum_{\lambda\in \Lambda} \phi_\lambda(x)\int_{\Rd}  f(y) \tilde \phi_\lambda(y)dy \quad \ {\rm for \ all} \  f\in V,
%  \end{equation}
%  and
%  \begin{equation}
%  g(x)=\sum_{\lambda\in \Lambda} \tilde \phi_\lambda(x) \int_{\Rd}g(y)  \phi_\lambda(y) dy \quad \ {\rm for \ all} \ g\in V^*.
%  \end{equation}

  \item[{(iv)}] Both $V$ and $V^*$ are   generated by $\Phi$ and $\tilde \Phi$ respectively in the sense that
 \begin{equation}\label{localizedframeforrange.tm.eq4}V=V_p(\Phi):=\Big\{ \sum_{\lambda\in \Lambda} c(\lambda) \phi_\lambda\Big|
 (c(\lambda))_{\lambda\in \Lambda}\in \ell^p(\Lambda)\Big\},\end{equation}
and
 \begin{equation}\label{localizedframeforrange.tm.eq5}V^*=V_{p/(p-1)}(\tilde \Phi):=\Big\{ \sum_{\lambda\in \Lambda} \tilde c(\lambda) \tilde \phi_\lambda\Big|
 (\tilde c(\lambda))_{\lambda\in \Lambda}\in \ell^{p/(p-1)}(\Lambda)\Big\}.\end{equation}
\end{itemize}
  \end{thm}

For  an orthogonal projection operator $T$ on $L^2(\Rd)$ whose kernel satisfies
 \eqref{kernel.assumption1} and \eqref{kernel.assumption2},
 we have the following result for its range space, which is established in \cite{sunacha08}
 under a weak assumption that
  the orthogonal projection  property for the operator $T$ is  replaced by the existence of a bounded pseudo-inverse.

\begin{cor}\label{localizedframe2.cr}
Let  $T$ be an orthogonal projection operator on $L^2(\Rd)$ whose integral
 kernel $K$  satisfies \eqref{kernel.assumption1} and \eqref{kernel.assumption2}, and
 let  $V$   be the  range space of the operator $T$ on $L^2(\Rd)$.
Then there exist a relatively-separated
subset $\Lambda$ and a family of functions $\Psi:=\{\psi_\lambda\}_{\lambda\in \Lambda}$ in $V$
such that
 $\Phi$  is  a localized  tight frame in the sense that
\begin{equation}\label{localizedframe2.cr.eq1}
\left\{\begin{array}
{l} |\psi_\lambda(x)|\le h(x-\lambda)\\
 \omega_\delta(\psi_\lambda)(x)\le h_\delta (x-\lambda)
 \quad \ {\rm for\ all} \ \lambda\in \Lambda \ {\rm and} \ x\in \Rd,\end{array}\right.
\end{equation}
where $h\in L^1(\Rd)$ and $h_\delta$ are integrable functions with
%\begin{equation}\label{localizedframe2.cr.eq2}
$\lim_{\delta\to 0} \|h_\delta\|_1=0$,  %\end{equation}
and
  \begin{equation}\label{localizedframe2.cr.eq3}
  f=\sum_{\lambda\in \Lambda} \langle f,  \psi_\lambda\rangle  \psi_\lambda \quad \ {\rm for \ all} \ f\in V.
  \end{equation}
Moreover,
 \begin{equation}\label{localizedframe2.cr.eq4}V=V_2(\Phi):=\Big\{ \sum_{\lambda\in \Lambda} c(\lambda) \psi_\lambda\Big|\
 (c(\lambda))_{\lambda\in \Lambda}\in \ell^2(\Gamma)\Big\}.\end{equation}
\end{cor}

\begin{rem}
The space $V_p(\Phi)$ was introduced in \cite{sunaicm08} to model signals with finite rate of innovations.
From Theorem \ref{localizedframeforrange.tm}, we see that signals in a reproducing kernel subspace  associated with
an idempotent operator  on $L^p(\Rd)$ with its kernel satisfying \eqref{kernel.assumption1} and \eqref{kernel.assumption2}
have finite rate of innovation.
\end{rem}

We conclude this subsection with  the proofs of Theorem \ref{localizedframeforrange.tm} and Corollary \ref{localizedframe2.cr}.

 \begin{proof}[Proof of Theorem \ref{localizedframeforrange.tm}]
%We follows the argument in \cite{sunacha08}.
Let $\delta_0>0$ be a sufficiently small positive number chosen later.
Define the operator $T_{\delta_0}$ by
\begin{equation} \label{localizedframeforrange.tm.newpf.eq1}
T_{\delta_0} f(x)=\int_{\Rd}  K_{\delta_0}(x, y) f(y) dy\quad \ f\in L^p(\Rd),
\end{equation}
where
\begin{equation} \label{localizedframeforrange.tm.newpf.eq2}
K_{\delta_0}(x,y)=\delta_0^{-d}\int_{[-\delta_0/2, \delta_0/2]^d} \int_{[-\delta_0/2, \delta_0/2]^d} \sum_{\lambda\in \delta_0\Zd} K(x, \lambda+z_1) K(\lambda+z_2, y) dz_1 dz_2.\end{equation}
Then
\begin{equation}\label{localizedframeforrange.tm.newpf.eq5}
T_{\delta_0} T=TT_{\delta_0}=T_{\delta_0}
\end{equation}
by \eqref{t2=t}, and
\begin{eqnarray}\label{localizedframeforrange.tm.newpf.eq6}\quad
   |K_{\delta_0}(x,y)-K(x,y)| %\\
%  &\le & \delta_0^{-d} \sum_{\lambda\in \delta_0\Zd} \int_{[-\delta_0/2, \delta_0/2]^d}
% \int_{[-\delta_0/2, \delta_0/2]^d}
% |K(x, \lambda+z_1)|  |K(\lambda+z_1,y)-K(\lambda+z_2,y)| dz_1 dz_2\\
 &\le &  \int_{\Rd} |K(x,z)| |\omega_{\delta_0}(K)(z, y)| dz
\end{eqnarray}
by Theorem \ref{kernel.tm}. Therefore
\begin{equation}\label{localizedframeforrange.tm.newpf.eq7}
\| T_{\delta_0}f-Tf\|_p\le r_1(\delta_0) \|f\|_p
\quad {\rm for \ all} \ f\in L^p,
\end{equation}
where
$$ r_1(\delta_0)=\big\|\sup_{z\in \Rd} |K(\cdot+z, z)|\big\|_1 \big\|\sup_{z\in \Rd} |\omega_{\delta_0}(K)(\cdot+z,z)\big\|_1.$$
Let $\delta_0>0$  be so chosen that $r_1(\delta_0)<1$. The existence of such a positive number follows from \eqref{kernel.assumption1}
and \eqref{kernel.assumption2}.
Then it follows from \eqref{localizedframeforrange.tm.newpf.eq5}, \eqref{localizedframeforrange.tm.newpf.eq6} and \eqref{localizedframeforrange.tm.newpf.eq7} that
 the operator  $T_{\delta_0}^\dag$ defined by
 \begin{equation}\label{localizedframeforrange.tm.newpf.eq8}
 T_{\delta_0}^\dag:=T+\sum_{n=1}^\infty (T-T_{\delta_0})^n
 \end{equation}
 is a bounded integral operator
 with the property that
 \begin{equation}\label{localizedframeforrange.tm.newpf.eq8bb}
 T_{\delta_0}^\dag T_{\delta_0}= T_{\delta_0} T_{\delta_0}^\dag=T,\end{equation}
 and that the kernel
 $K_{D,\delta_0}$ of the operator  $T_{\delta_0}^\dag$ satisfies
 \begin{equation}\label{localizedframeforrange.tm.newpf.eq9}
 \big\|\sup_{z\in \Rd} |K_{D, \delta_0}(\cdot+z, z)|\big\|_1<\infty
 \end{equation}
 and
 \begin{equation} \label{localizedframeforrange.tm.newpf.eq10}
\lim_{\delta\to 0} \big\|\sup_{z\in \Rd} |\omega_\delta(K_{D, \delta_0})(\cdot+z, z)|\big\|_1=0.
 \end{equation}

Define
\begin{equation} \label{localizedframeforrange.tm.newpf.eq11}
\left\{\begin{array}{l}
\phi_\lambda(x)=\delta_0^{-d/p} \int_{\Rd}  \int_{[-\delta_0/2, \delta_0/2]^d} K_{D, \delta_0}(x, z_1) K(z_1, \lambda+z_2) dz_2 dz_1
\\
\tilde \phi_\lambda(x) =\delta_0^{-d+d/p} \int_{[-\delta_0/2, \delta_0/2]^d}
K(\lambda+z,  x) dz \
\end{array}\right.\end{equation}
 for all $\lambda\in \delta_0\Zd$,
% \ {\rm and} \ \phi_\lambda=T_{\delta_0}^\dag \psi_\lambda, \ \lambda\in \delta_0\Zd.
 and set
 $$\Phi=\{\phi_\lambda\}_{\lambda\in \delta_0\Zd}\ \ {\rm  and}  \ \ \tilde \Phi=\{\tilde \phi_\lambda\}_{\lambda\in \delta_0\Zd}.$$
Now let us verify that the above two families $\Phi$ and $\tilde \Phi$ of functions  satisfy all required properties.
By \eqref{localizedframeforrange.tm.newpf.eq8},
\eqref{localizedframeforrange.tm.newpf.eq11} and Theorem \ref{kernel.tm},
\begin{equation}\label{localizedframeforrange.tm.newpf.eq12}\phi_\lambda\in V\ {\rm and}
\ \tilde \phi_\lambda\in V^* \ {\rm  for \ all} \ \lambda\in \delta_0\Zd.
\end{equation}
Moreover,
\begin{eqnarray*}
 |\phi_\lambda(x)|  & \le &   \delta_0^{-d/p} \int_{[-\delta_0, \delta_0]^d} \int_{\Rd}
 \big(\sup_{z'\in \Rd} |K(y-z+z', z')|\big)\\
 & & \quad  \times \big(\sup_{z'\in \Rd}|K_{D, \delta_0}(x-\lambda-y+z', z')|\big)
 dy dz,\nonumber\\
%\end{eqnarray*}
%\begin{eqnarray*}
 |\omega_{\delta} (\phi_\lambda)(x)|  &  \le  &  \delta_0^{-d/p} \int_{[-\delta_0, \delta_0]^d} \int_{\Rd}
\big(\sup_{z'\in \Rd} |K(y-z+z', z')|\big) \\
 & & \quad \times \big(\sup_{z'\in \Rd}|\omega_\delta(K_{D, \delta_0})(x-\lambda-y+z', z')|\big)
 dy dz,\nonumber
\end{eqnarray*}
\begin{equation*}
 |\tilde \phi_\lambda(x)|   \le   \delta_0^{-d(p-1)/p} \int_{[-\delta_0, \delta_0]^d}
\big(\sup_{z'\in \Rd} |K(\lambda-x+z+z', z')|\big) dz,
\end{equation*}
and
\begin{equation*}
 |\omega_\delta(\tilde \phi_\lambda)(x)|  \le  \delta_0^{-d(p-1)/p} \int_{[-\delta_0, \delta_0]^d}
\big(\sup_{z'\in \Rd} |\omega_\delta(K)(\lambda-x+z+z', z')|\big) dz.
\end{equation*}
The above estimates for $\phi_\lambda, \omega_\delta(\phi_\lambda), \tilde \phi_\lambda$ and $\omega_{\delta}(\tilde \phi_\lambda)$,
together with \eqref{kernel.assumption1}, \eqref{kernel.assumption2}, \eqref{localizedframeforrange.tm.newpf.eq9}
and \eqref{localizedframeforrange.tm.newpf.eq10}, prove
\eqref{localizedframeforrange.tm.eq2} and hence the localized frame property (i) for $\Phi$ and $\tilde \Phi$.

By \eqref{localizedframeforrange.tm.newpf.eq8bb}, we have
\begin{equation}
f=T_{\delta_0}^\dag T_{\delta_0} f=
\sum_{\lambda\in \Lambda} \langle f, \tilde \phi_\lambda\rangle \phi_\lambda \quad \ f\in V,
\end{equation}
and
\begin{equation}
g=T_{\delta_0}^* (T_{\delta_0}^\dag)^* g=
%\sum_{\lambda\in \Lambda} \langle (T_{\delta_0}^\dag)^* g, \psi_\lambda\rangle \tilde \psi_\lambda=
\sum_{\lambda\in \Lambda} \langle g, \phi_\lambda\rangle \tilde \phi_\lambda\quad \ {\rm for \ all} \ g\in V^*.
\end{equation}
Then the dual pair property  (ii) for $\Phi$ and $\tilde \Phi$ follows.

The $p$-frame property for $\tilde \Phi$ and the $p/(p-1)$-frame property for $\Phi$ follow from
the localization property (i) and
the dual pair property  (ii)
for $\Phi$ and $\tilde \Phi$.  We leave the detailed proof for the interested readers.

The inclusion $V\subset V_p(\Phi)$ follows from the $p$-frame property
for $\tilde \Phi$ and the reconstruction formula
$f=\sum_{\lambda\in \Lambda} \langle f, \tilde \phi_\lambda\rangle \phi_\lambda$
for any $f\in V$. The reverse inclusion $V_p(\Phi)\subset V$ follows from
\eqref{localizedframeforrange.tm.newpf.eq12} and
the closedness of the space  $V$ in $L^p$. This proves that $V=V_p(\Phi)$.
The conclusion $V^*=V_{p/(p-1)}(\tilde \Phi)$ can be established by similar arguments.
\end{proof}

\begin{proof}[Proof of Corollary \ref{localizedframe2.cr}]
By the argument in the proof of  Theorem \ref{localizedframeforrange.tm}, there exists a positive
number $\delta_0$  and a family $\Phi=\{\phi_\lambda\}_{\lambda\in \delta_0\Zd}$ of functions
in $V$  such that $\Phi$ is a localized frame for $V$.
Then
 the autocorrelation matrix
 %\begin{equation} \label{localizedframeforrange.tm.pf.eq7}
 $A_{\Phi, \Phi}:=(\langle \phi_{\lambda}, \phi_{\lambda'}\rangle)_{\lambda, \lambda'\in\delta_0\Zd}$ %\end{equation}
 belongs to the Gohberg-Baskakov-Sj\"ostrand class  ${\mathcal C}(\delta_0\Zd)$
 by the localization property \eqref{localizedframeforrange.tm.eq2}. Here
 the Gohberg-Baskakov-Sj\"ostrand class  ${\mathcal C}(\delta_0\Zd)$ is given by
   \begin{eqnarray}\label{localizedframeforrange.tm.pf.eq9}
 {\mathcal C}(\delta_0\Zd) & = &
 \Big\{A:= (a(\lambda, \lambda'))_{\lambda,\lambda'\in \delta_0\Zd}\ \Big| \ \\
 & & \quad \  \|A\|_{\mathcal C}:=\sum_{\lambda\in \delta_0\Zd} \sup_{\lambda'\in \delta_0\Zd}
|a(\lambda+\lambda', \lambda')|<\infty\Big\} \nonumber\end{eqnarray}
\cite{baskakov,  gohberg89, ssjfa09, sjostrand2, suntams}.
By the frame property for $\Phi$,
the square root of the  autocorrelation matrix $A_{\Phi, \Phi}$ has bounded
    pseudo-inverse $(A_{\Phi, \Phi}^{1/2})^\dag$. %:=(b(\lambda, \lambda'))_{\lambda, \lambda'\in \delta_0\Zd}$.
This together with
 the  Wiener's lemma
for infinite matrices in  the Gohberg-Baskakov-Sj\"ostrand class
\cite{baskakov,  gohberg89,  ssjfa09, sjostrand2, suntams} shows that
 $$(A_{\Phi, \Phi}^{1/2})^\dag:=(b(\lambda, \lambda'))_{\lambda, \lambda'\in \delta_0\Zd}\in {\mathcal C}(\delta_0\Zd).$$
Then one may easily verify that
$\Psi=\{\psi_\lambda\}_{\lambda\in \delta_0\Zd}$ is a  localized tight frame  for $V$ that has all the  required properties
\eqref{localizedframe2.cr.eq1}--\eqref{localizedframe2.cr.eq4}, where
$$\psi_\lambda=\sum_{\lambda'\in \delta_0\Zd} b(\lambda, \lambda') \phi_{\lambda'}, \ \lambda\in \delta_0\Zd.$$
\end{proof}

\subsection{Examples}
In this subsection, we present two examples of a reproducing kernel space associated with
an idempotent integral operator on $L^p$.

\begin{ex}\label{modelspace1.example} {\rm \cite{schumakerbook}\
Let $n\ge 1$, $\Lambda=\{\lambda_k\}_{k\in \ZZ}$ be  a bi-infinite increasing sequence of real numbers with
$$0<\inf_{k\in \ZZ} (\lambda_{k+1}-\lambda_k)\le \sup_{k\in \ZZ} (\lambda_{k+1}-\lambda_k)<\infty,$$
and \begin{eqnarray}
S_n^{n-1}(\Lambda) & = & \Big\{ f\in C^{n-1}(\RR): \ \ f|_{[\lambda_k,\lambda_{k+1}]}\  {\rm is \ a \ polynomial\ }\ \\
 & &  \quad {\rm having\ degree \ at\ most} \  n  \  {\rm for \ each}\ k\in \ZZ\Big\}.\nonumber\end{eqnarray}
Let $B_i$ be the normalized B-spline  associated
with the knots $\lambda_i, \ldots, \lambda_{i+n+1}$, and define its autocorrelation matrix $A=\big(\langle B_i, B_j\rangle\big)_{i,j\in \ZZ}$. Then
the infinite matrix $A$  is invertible and its inverse $B=(b_{ij})_{i,j\in \ZZ}$ has exponential off-diagonal decay, that is,
there exist constants $C$ and $\epsilon$ such that
$$|b_{ij}|\le C \exp(-\epsilon |i-j|)\quad \ i,j\in \ZZ.$$
Define
$$K(x,y)=\sum_{i,j\in \ZZ} B_i(x) b_{ij} B_j(y)$$
and
$$Tf(x)=\int_{\RR} K(x,y) f(y)dy.$$
Then one may verify that  the above integral operator $T$ is an idempotent operator on $L^p(\RR)$,
the  kernel $K$ of the operator $T$ satisfies \eqref{kernel.assumption1} and \eqref{kernel.assumption2},
and $S_n^{n-1}(\Lambda)\cap L^p(\RR)$ is the range of the operator $T$ on $L^p(\RR)$.
The spline model has many practical advantages over the  band-limited model in Shannon's sampling theory,
and has been well-studied (see \cite{sunsiam, uieee00, wbook90} and the  references therein).
}
\end{ex}

\begin{ex}\label{modelspace2.example} {\rm \cite{sunaicm08}\
Let $\Lambda$ be a relatively-separated subset of $\Rd$ with positive gap,
$\Phi=\{\phi_\lambda\}_{\lambda\in \Lambda}$  and $\tilde \Phi=\{\tilde \phi_\lambda\}_{\lambda\in \Lambda}$
be two  families of functions such that
$$|\phi_\lambda(x)|+|\tilde \phi_\lambda(x)|\le h(x-\lambda), \ x\in \Rd,$$
and
$$|\omega_\delta(\phi_\lambda)(x)|+|\omega_\delta(\tilde \phi_\lambda)(x)|\le h_\delta (x-\lambda),\  x\in \Rd,$$
hold for all $\lambda\in \Lambda$ and $\delta>0$, where $h$ and $h_\delta$  are functions in the Wiener amalgam space
${\mathcal W}$
with $\lim_{\delta\to 0} \|h_\delta\|_{\mathcal W}=0$. %, see \eqref{wieneramalgam.def}.
Then one may verify that
the kernel function
\begin{equation}\label{kerneltype1}
K(x,y):=\sum_{\lambda\in \Lambda} \phi_\lambda(x) \tilde \phi_{\lambda}(y)\end{equation}
satisfies \eqref{kernel.assumption1} and \eqref{kernel.assumption2}.
If  we further assume that $\Phi$ and $\tilde \Phi$  satisfy
$$ \int_{\Rd} \phi_\lambda(x) \tilde \phi_{\lambda'}(x)dx=\delta_{\lambda,\lambda'} \ {\rm for \ all} \ \lambda, \lambda'\in \Lambda,$$
where $\delta_{\lambda, \lambda'}$ stands for the Kronecker symbol,
then the operator $T$ with the kernel $K$ in \eqref{kerneltype1} is an idempotent operator on $L^2$. In this case,
\begin{equation}\label{lfg_rks.def}
V_2(\Phi):=\Big\{\sum_{\lambda\in \Lambda} c(\lambda) \phi_\lambda(x)\ \big|\ \ \sum_{\lambda\in \Lambda} |c(\lambda)|^2<\infty\Big\}
\end{equation}
is the range space of the operator $T$ on $L^2$ and hence a reproducing kernel subspace of $L^2$.
A special case of the above space $V_2(\Phi)$
is the  finitely-generated shift-invariant space $V_2(\phi_1, \ldots, \phi_r)$ in \eqref{sis.def}, see \cite{abk09, agsiam01, astjfaa05, lwjfaa96} and references therein.
}\end{ex}

\end{appendix}

{\bf Acknowledgement}: \ The authors would like to thank Professors Akram Aldroubi,
Deguang Han, Karlheinz Gr\"ochenig,  Wai-Shing Tang, and Jun Xian for their discussion and suggestions.

\end{document}